\PassOptionsToClass{12pt}{revtex4-1}
\documentclass[preprint]{aastex631}
\usepackage{stix2}
\usepackage{anyfontsize}
\usepackage{amsmath}
\usepackage{rotating, multirow, ctable,xcolor,isomath}

\graphicspath{{./}{./images/}}

  \newcommand\eps{\epsilon}
\newcommand{\lapprox} {\, \lower3pt\hbox{$\sim$}\llap{\raise2pt\hbox{$<$}}\,}
\newcommand{\gapprox} {\, \lower3pt\hbox{$\sim$}\llap{\raise2pt\hbox{$>$}}\,}

\newcommand\mmatrix[1]{ \mathsfbfit{{#1}}} 
\renewcommand{\vec}[1]{\mathbf{#1} }

\shorttitle{Plasma motions and compressive wave energetics in the solar corona}
\shortauthors{Azzollini et al.}

\begin{document}

\title{Plasma motions and compressive wave energetics in the solar corona and solar wind from radio wave scattering observations}

\author[0009-0001-7368-0938]{Francesco Azzollini}
\affiliation{School of Physics \& Astronomy, University of Glasgow, Glasgow, G12 8QQ, UK}

\author[0000-0001-8720-0723]{A. Gordon Emslie}
\affiliation{Department of Physics \& Astronomy, Western Kentucky University, Bowling Green, KY 42101, USA}

\author[0000-0003-1967-5078]{Daniel L. Clarkson}
\affiliation{School of Physics \& Astronomy, University of Glasgow, Glasgow, G12 8QQ, UK}

\author[0000-0002-4389-5540]{Nicolina Chrysaphi}
\affiliation{Sorbonne Universit\'{e}, \'{E}cole Polytechnique, Institut
Polytechnique de Paris, CNRS, Laboratoire de Physique des Plasmas (LPP), 75005
Paris, France}
\affiliation{School of Physics \& Astronomy, University of Glasgow, Glasgow, G12 8QQ, UK}

\author[0000-0002-8078-0902]{Eduard P. Kontar}
\affiliation{School of Physics \& Astronomy, University of Glasgow, Glasgow, G12 8QQ, UK}

\begin{abstract}

Radio signals propagating via the solar corona and solar wind are significantly
affected by compressive waves, 
impacting properties of solar bursts as well as sources viewed through 
the turbulent solar atmosphere. While static fluctuations
scatter radio waves elastically, moving, turbulent or oscillating density
irregularities act to broaden the frequency of the scattered waves. 
Using a new anisotropic density fluctuation model 
from the kinetic scattering theory for solar radio bursts, 
we deduce the plasma velocities required to explain observations of spacecraft signal
frequency broadening. The inferred velocities are consistent with motions that
are dominated by the solar wind at distances $\gapprox 10 \, R_\odot$, but the
levels of frequency broadening for $\lapprox 10 \, R_\odot$ require additional
radial speeds $\sim$(100-300)~km~s$^{-1}$ and/or transverse speeds
$\sim$(20-70)~km~s$^{-1}$. The inferred radial velocities also appear consistent with
the sound or proton thermal speeds, while the speeds perpendicular to the radial
direction are consistent with non-thermal motions measured via coronal
Doppler-line broadening, interpreted as Alfv\'enic fluctuations. Landau damping
of parallel propagating ion-sound (slow MHD) waves allow an estimate of the
proton heating rate. The energy deposition rates due to ion-sound
wave damping peak at a heliocentric distance of $\sim$$(1-3) \, R_\odot$ are
comparable to the rates available from a turbulent cascade of Alfv\'enic waves
at large scales, suggesting a coherent picture of energy transfer, via the
cascade or/and parametric decay of Alfv\'en waves to the small scales where
heating takes place.
\end{abstract}


\section{Introduction}

Radio waves propagating in the solar atmosphere are scattered by density
fluctuations, affecting the observed properties of radio emission originating
either in the solar corona itself or in distant sources close to the location of
the Sun in the sky. Solar burst source angular sizes and time profiles are
broadened by the density fluctuations, and the apparent source positions are shifted, 
 usually away from the Sun.
Typically, extrasolar sources are observed at frequencies $\omega$ that are significantly higher than the plasma frequency $\omega_{p e}$ of the solar medium through which their emitted radiation 
propagates and scatters. This results in a scattering mean free path 
that is much larger than the distance traveled by the radio waves 
in the turbulent medium. 
Consequently, such sources experience much weaker scattering than solar 
radio burst emission at frequencies inherently close to the local plasma
frequency (or its second harmonic). 
Despite this distinction, 
the properties of solar and extrasolar sources are affected by the same
density turbulence, with recent study showing 
that the propagation of solar and extrasolar radio emission in both strong and weak scattering regimes can be modeled using the same kinetic approach
\citep{2023ApJ...956..112K}. 
Since the speeds of density fluctuations are much less than the speed
of radio waves, scattering is often treated as elastic, resulting primarily in
angular broadening of extra-solar point sources
\cite[e.g.,][]{1952Natur.170..319M,1958MNRAS.118..534H,1972PASA....2...84B,1994JApA...15..387A,2015MNRAS.447.3486I},
an increase in the angular source sizes of solar radio sources
\cite[e.g.,][]{1965BAN....18..111F,1972A&A....18..382S,1974SoPh...35..153R,
1994ApJ...426..774B,1999A&A...351.1165A,2008ApJ...676.1338T,2017NatCo...8.1515K,
2018ApJ...868...79C,2018ApJ...857...82K,2018SoPh..293..132M,
2019ApJ...873...48G,2019ApJ...884..122K,2020ApJ...893..115C,
2020ApJS..246...57K,2021A&A...645A..11M,2021ApJ...909....2M,
2021A&A...655A..77M,2021A&A...656A..34M,2021ApJ...913..153S,2021ApJ...917L..32C}
and changes in the paths of spacecraft signals observed through the turbulent
solar atmosphere
\citep[e.g.,][]{1978ApJ...219..727W,1980R&QE...22..728R,1980MNRAS.190P..73B,
1982SSRv...33...99B,1994GeoRL..21...85W,2013CosRe..51...13E,2018AdSpR..61..552Y,
2019ApJ...871..202W,2020CosRe..58..460E,2022SoPh..297...34C}. The density
fluctuations are also routinely observed near 1 au
\citep[e.g.,][]{1983A&A...126..293C,1990JGR....9511945M,
2020ApJS..246...57K,2024arXiv240205191W}.

The collective Compton scattering from electron density fluctuations 
\citep[e.g.,][]{1958JETP....6..576A,1960RSPSA.259...79D,2003PhPl...10.3297B} that are
moving or oscillating perpendicular to the direction of wave propagation can,
however, lead to an \emph{inelastic} change in the wavenumber and hence
frequency broadening (that is normally a small fraction of the observed
frequency). Doppler broadening of radio waves from spacecraft has been
extensively studied to diagnose expansion of the solar wind
\citep[e.g.,][]{1978ApJ...219..727W}, and moving density irregularities have
also been analyzed via observations of interplanetary scintillations
\cite[e.g.,][]{1993Natur.366..543W,1996Ap&SS.243...97W,2009RaSc...44.6004M,2015SoPh..290.2539M}.
The spectral width of the radio-wave signal, or the strength of scintillation,
is proportional to the speed of the density irregularity weighted by the
amplitude of the density fluctuation. Using multiple receivers to observe
interplanetary scintillation, \citet{1971A&A....10..310E} found a random
velocity component of $v \simeq$~(100-200)~km~s$^{-1}$ at (5-10)~$R_\odot$ and
less than 50~km~s$^{-1}$ at 40~$R_\odot$. Somewhat lower fractional velocities
$\delta v/v \simeq 0.25$ of the solar wind speed were deduced by
\cite{1972JGR....77.4602A}. \citet{1986ASSL..123...59A} reported a random
velocity component at $<12 \, R_\odot$ that was comparable to the bulk flow
speed. Assuming a density model of the corona and  the fractional amplitude of
density fluctuations, \cite{2020SoPh..295..111W} used spacecraft carrier
frequency fluctuations to infer the flow velocity profile in the middle 
corona \citep{2023SoPh..298...78W}.

Here we use a recently-developed anisotropic density turbulence model
\citep{2023ApJ...956..112K} to analyze a large observational dataset of Doppler
broadening of spacecraft carrier frequencies and thus determine the speeds of
density fluctuations in the space between the Sun and 1~au. The spectral
broadening is discussed in terms of solar wind flows, compressive waves, and
random plasma motions. By matching observations, we determine the characteristic
velocities of density fluctuations and we show how the wavevector anisotropy,
$\alpha = q_\parallel/q_\perp$, associated with density fluctuations along
versus perpendicular to the solar radius vector, affects these results. The
average frequency broadening at $>$10~$R_\odot$ is found, in line with the
previous works, to be determined mostly by the radial solar wind speed, while
closer to the Sun ($<$10~$R_\odot$), both transverse and radial motions could
contribute. Due to the wavenumber anisotropy in the density fluctuations, which
are typically elongated along the radial direction $(\alpha < 1)$, smaller
perpendicular velocities are needed to explain a given amount of frequency
broadening. For example, if $\alpha =0.25$, either radial speeds
$\simeq$$160$~km~s$^{-1}$ or transverse speeds $\simeq$$40$~km~s$^{-1}$ are
consistent with the observed amount of frequency broadening.

In Section~\ref{sec:static-fluctuations} we introduce the model used. In
Section~\ref{sec:inelastic} we consider the effects of inelastic scattering in
the presence of motions perpendicular to the line of sight, and we derive the
associated diffusion tensor for radio waves in anisotropic turbulent plasma. We
also compare the deduced random motions with non-thermal velocities deduced from
the emission line broadening of hot ions.

In Sections~\ref{sec:observations} and~\ref{sec:freq_broad_measurements} we turn
our attention to observed values of the frequency broadening $\Delta f$. We
derive an expression for the fractional frequency broadening $\Delta f/f$ in
terms of the product of the fluid velocity perpendicular to the line of sight
and the wavenumber-weighted line-of-sight integral of the density fluctuation
amplitude. Using a density fluctuation model (deduced from observations of solar
radio bursts and angular broadening of extra-solar sources) we then deduce the
magnitudes of the flow velocities (and/or phase speeds) required to produce the
observed frequency broadenings. The inferred speeds are compared with other
characteristic speeds, such as the sound speed, the Alfv\'en speed, and the
solar wind speed.

Sections~\ref{sec:alfven_cascade}
and~\ref{sec:ion_sound_wave_damping} compare the power supplied by large-scale
Alfv\'enic kink/shearing motions with the heating effected by damping of the
ion-sound waves associated with density fluctuations at much shorter length
scales. In Section~\ref{sec:summary} we summarize the results obtained and
present our conclusions. A number of Appendices present some of the more tedious
mathematical derivations of certain results.

\section{Static density fluctuations and angular
broadening}\label{sec:static-fluctuations}

Density fluctuations with wavevector $\mathbf{q}$ are characterized by their
three-dimensional wavevector spectrum $S(\vec{q})$, typically normalized \citep[see, e.g.,][]{2023ApJ...956..112K} by the rms level of fluctuations in the local density $n$ (cm$^{-3}$):

\begin{equation}\label{s-q-def}
\int S(\vec{q}) \, \frac{d^3 q}{(2 \pi )^3} = \frac{\langle \delta n^2 \rangle}{n^2} \equiv \epsilon^2 \,\,\, .
\end{equation}
Following previous studies
\citep[e.g.,][]{1999A&A...351.1165A,2019ApJ...873...33B,2019ApJ...884..122K},
the diffusion tensor describing elastic scattering of radio waves with
wavevector $\vec{k}$ in a medium containing static density fluctuations can be
written as

\begin{equation}\label{eq:D}
  D_{i j}=\frac{\pi \omega_{p e}^4}{4 \, \omega^2} \int q_i \, q_j \, S(\vec{q}) \, \delta\left(\vec{q} \cdot \vec{v}_g\right) \, \frac{d^3 q}{(2 \pi)^3} \,\,\, ,
\end{equation}
where $q_i, q_j$ (cm$^{-1}$) are the components of the density fluctuation
wavevector in the directions labeled by the suffices $i,j$, the wave group
velocity $\vec{v}_g=\partial \omega/\partial \vec{k}$ and $\omega
(\vec{k})=(\omega_{pe}^2 + c^2 \, k^2)^{1/2}$ is the angular frequency of
electromagnetic waves with wavevector $\vec{k}$ in a plasma with local plasma
frequency $\omega_{pe} (\vec{r})$.

Similar to \cite{2023ApJ...956..112K}, we take the spectrum of density
turbulence to be anisotropic with a constant anisotropy factor $\alpha$, so that

\begin{equation}\label{eq:Sq_alpha}
  S(\vec{q})=S(\tilde{q})\,,\,\,\, \mbox{where}\;\;\;
  \tilde{q}=\sqrt{\frac{q_{\parallel}^2}{\alpha^2} + q_{\perp_2}^2+ q_{\perp_1}^2} \,\,\, ,
\end{equation}
which has axial symmetry around the $\parallel$ direction, i.e., along the
magnetic field $\vec{B}$, and is isotropic with respect to the $\vec{\Tilde q}$
basis. In matrix form, $\Tilde{\vec{q}} = \mmatrix{A} \, \vec{q} = (\alpha^{-1}
q_\parallel, q_{\perp_2}, q_{\perp_1})$, where $\mmatrix{A}$ is the anisotropy
matrix

\begin{equation}\label{eq:matrixA}
\mmatrix{A}=\left(\begin{array}{ccc}
\alpha^{-1}&0&0\\
0&1&0\\
0&0&1
\end{array}\right) \,\,\, . 
\end{equation}

The quantity $\alpha$ appearing in Equations~\eqref{eq:Sq_alpha}
and~\eqref{eq:matrixA} quantifies the degree of anisotropy in the turbulence
distribution: $\alpha <1$ corresponds to density fluctuations elongated along
the magnetic field  ($q_\parallel^{-1} > q_\perp^{-1}$, i.e., $q_\parallel <
q_\perp$), as is often observed in the solar wind
\cite[e.g.,][]{1987A&A...181..138C,2021A&A...656A..34M}. Typical reported values
of $\alpha$ are $0.1-0.4$
\citep[e.g.,][]{1972PASAu...2...86D,1989ApJ...337.1023C,1990ApJ...358..685A,2019ApJ...884..122K,2020ApJ...905...43C,2020ApJ...898...94K,2023ApJ...946...33C}.

For a radio wave propagating with $\vec{v}_g \simeq c \, \vec{k}/{\vert
\vec{k}\vert}$ along the $\perp_1$-direction (see Figure~\ref{fig:coords}), we
can find the components of the diffusion tensor elements, viz.
(Equation~\eqref{eq:Delastic})

\begin{equation}\label{eq:D_ellastic}
\mmatrix{D}= 
\frac{\pi \omega_{p e}^4}{16 \, \omega^2 c} \, \overline{q \, \eps^2} \,
\left(\begin{array}{ccc}
\alpha^2&0&0\\
0&1&0\\
0&0&0
\end{array}\right ) 
\,\,\, ,
\end{equation}
where we have introduced the spectrum-weighted mean wavenumber $\overline{q
\,\eps^2}$, defined as \citep{2023ApJ...956..112K}

\begin{equation}\label{q-eps2-definition}
  \overline{q \, \eps ^2} = \int q \, S(q) \, \frac{d^3 q}{(2\pi)^3} =\alpha \int \Tilde q \, S({\Tilde q}) \, \frac{d^3\Tilde{q}}{(2\pi)^3} =  \alpha \, \frac{4 \pi}{(2 \pi)^3} \, \int {\Tilde q}^3 \, S({\Tilde q}) \, d\Tilde q  \,\,\, .
\end{equation}
where in the second equality, we have transformed from the ${\bf q}$ basis to the
$\Tilde{\bf q}$ basis, in which the wavenumber spectrum is isotropic.

The diffusion tensor $\mmatrix{D}$ given by Equation~\eqref{eq:D_ellastic} has
two non-zero elements that determine the scattering rates $d\langle\Delta
k_\parallel^2\rangle/dt$ and $d\langle\Delta k_{\perp_2}^2\rangle/dt$. The
$\perp_2$ and $\parallel$ directions are perpendicular to the wave propagation
vector ${\bf k}$ (see Figure \ref{fig:coords} in Appendix~\ref{C:static}). 
For $\Delta \mathbf{k} \perp \mathbf{k}$, the scattering
corresponding to terms $d\langle\Delta k_\parallel^2\rangle/dt$ and
$d\langle\Delta k_{\perp_2}^2\rangle/dt$ is elastic: $|\mathbf{k}+\Delta
\mathbf{k}|^2 \simeq|\mathbf{k}|^2+2 \mathbf{k} \cdot \Delta
\mathbf{k}=|\mathbf{k}|^2$, i.e., $|\mathbf{k}|$ is a constant. To change the
absolute value of $|\mathbf{k}|$, or equivalently the wave frequency
$\omega(\mathbf{k}) \simeq c|\mathbf{k}|$, we must have non-zero $\mathbf{k}
\cdot \Delta \mathbf{k}$, i.e. $d\langle\Delta k_{\perp_1}^2\rangle / d t \neq
0$. The effects of such inelastic scatterings, which are important for frequency
broadening, are next considered.

\section{Inelastic scattering of radio waves}\label{sec:inelastic}

When the density fluctuations are not static, but are instead due to either
waves or density fluctuations advected by plasma motions, the scattering could
be inherently inelastic with $d\langle \Delta k_{\perp_1}^2\rangle/dt\neq 0$, so
that $|{\vec{k}}|\neq \mbox{const}$, leading to a change in the wave frequency
$\omega$. Consider an electromagnetic (EM) wave with energy $\omega(\vec{k})$
and wavevector $\vec{k}$ that is scattered by a density fluctuation with
wavenumber $\vec{q}$ and frequency $\Omega (\vec{q})$, resulting in a scattered
EM wave with frequency $\omega(\vec{k}')$ and wavevector $\vec{k}'$. Momentum
and energy conservation in such a three-wave process \citep{1995lnlp.book.....T}
demands that

\begin{equation}\label{eq:momentum_energy_conservations}
    \vec{k} + \vec{q} = \vec{k}', \;\;\; \omega(\vec{k}) + \Omega (\vec{q}) = \omega(\vec{k}') \,\,\, .
\end{equation}
Using the dispersion relation for electromagnetic waves $\omega^2(\vec{k}) =
\omega_{pe}^2+c^2k^2$ and using the resonance condition $\Omega(\vec{q}) =
\vec{v} \cdot \vec{q}$ as a dispersion relation, one finds that, for $\omega \gg
\omega_{pe}$, $|\vec{k}| \gg |\vec{q}|$, and $|\vec{k}'| \gg |\vec{q}|$,

\begin{equation}
\vec{q} \cdot \frac{\vec{k}}{|\vec{k}|} \simeq \frac{\Omega (\vec{q})}{c} =\frac{\vec{q} \cdot\vec{v}}{c} \,\,\, .
\end{equation}
Hence to satisfy the conservation of energy and momentum
relations~(\ref{eq:momentum_energy_conservations}), the density wavevector
$\vec{q}$ should be quasi-perpendicular to $\vec{k}$:

\begin{equation}\label{eq:vperpk}
\frac{q_{\parallel \vec{k}}}{q_{\perp \vec{k}}} \simeq \frac{v_{\perp \vec{k}}}{c} \ll 1 \,\,\, ,
\end{equation}
showing that density fluctuations involving motions perpendicular to $\vec{k}$
($v_{\perp \vec{k}} \neq 0$) produce a shift in the the magnitude $|\vec{k}|$ of
the electromagnetic wavevector and hence in the frequency of the electromagnetic
wave. 

\subsection{Radially moving density fluctuations}\label{sec:radial}

The generalization of Equation~\eqref{eq:D} in the presence of non-static
density fluctuations is

\begin{equation}\label{eq:Dmoving}
  D_{i j}=\frac{\pi \omega_{p e}^4}{4 \, \omega^2} \int q_i \, q_j \, S(\vec{q}) \,
  \delta\left(\Omega (\vec{q})- \vec{q} \cdot \vec{v}_g\right) \, \frac{d^3 q}{(2 \pi)^3} \,\,\, ,
\end{equation}
where $\Omega(\vec{q})$ is the dispersion relation for the density fluctuations,
which is conceptually identical to plasma wave scattering on plasma density
fluctuations
\citep{1969npt..book.....S,1982PhFl...25.1062G,2012ApJ...761..176R}.

We first consider density fluctuations moving along the radial direction
$\vec{B}$, i.e., $\Omega (\vec{q}) = v_\parallel \, q_\parallel$. In this case,
the change of frequency  (or absolute value of the wavevector of the radio wave)
is due to non-zero values of $v_\parallel$. One can include the effects of the
moving waves in the calculation of the components $D_{ij}$ of the diffusion
tensor that affects both the direction of propagation (angular broadening) and
change in wavenumber (frequency broadening). As shown in
Equation~\eqref{eq:D-parallel-motions},
the modified diffusion tensor takes the form

\begin{equation}\label{eq:D_vparallel}
\mmatrix{D}=
\frac{\pi \omega_{p e}^4}{16 \, \omega^2 \, c} \, \overline{q \, \epsilon^2} \,
\left(\begin{array}{ccc}
\alpha^2&0&0\\
0&1&0\\
0&0&\frac{\alpha^2 v_{\parallel}^2}{c^2}
\end{array}\right) \,\,\, .
\end{equation}
Naturally, Equation~\eqref{eq:D_vparallel} reduces to  
Equation~\eqref{eq:D_ellastic} when $v_\parallel \rightarrow 0$.

\subsection{Transverse density fluctuations}\label{sec:perp}

We can similarly evaluate the diffusion tensor components $D_{ij}$ for the case
of waves moving in the perpendicular (transverse) direction, with assumed
dispersion relation $\Omega (\vec{q}) = v_{\perp_2} q_{\perp_2}$. 
Substituting this into Equation~\eqref{eq:Dmoving}
gives the form of the diffusion tensor (see Equation~\eqref{eq:D:perp-motions})

\begin{equation}\label{eq:D_vperp}
\mmatrix{D}=\frac{\pi \omega_{p e}^4}{16 \, \omega^2 \, c} \, \overline{q \, \epsilon^2} \,
\left(\begin{array}{ccc}
\alpha^2&0&0\\
0&1&0\\
0&0&\frac{{v}_{\perp_2}^2}{c^2}
\end{array}\right) \,\,\, ,
\end{equation}

Equations~\eqref{eq:D_vparallel} and~\eqref{eq:D_vperp} show that all motions in
the plane of the sky, i.e., perpendicular to the radio wave propagation
direction, lead to a change in the absolute magnitude of the radio-wave
wavenumber. If there are waves in both the parallel ($\parallel$) and
perpendicular ($\perp_2$) directions, the diffusion effects simply add together:

\begin{equation}\label{eq:D_vperp_vparallel}
\mmatrix{D}=\frac{\pi \omega_{p e}^4}{16 \, \omega^2 \, c} \, \overline{q \, \epsilon^2} \,
\left(\begin{array}{ccc}
\alpha^2&0&0\\
0&1&0\\
0&0&\frac{\alpha^2 v_{\parallel}^2+v_{\perp_2}^2}{c^2}
\end{array}\right) \,\,\, .
\end{equation}
We note that since $\alpha <1$ perpendicular motions are more effective at
frequency broadening than radial (parallel) motions. 

\subsection{Random (turbulent) motions}\label{sec:random}

Observation of solar corona UV spectral lines often show significant broadening in excess
of the thermal width at which the responsible atomic species is formed
\citep[e.g.,][]{1990ApJ...348L..77H,1991SoPh..131...25C,1998A&A...339..208B,1998SoPh..181...91D,1999A&A...349..956D,1999ApJ...510L..63E,2004AnGeo..22.3055C,2009A&A...501L..15B,2009ApJ...691..794L,2011SoPh..270..213S}.
Such non-thermal broadening of lines is normally interpreted as the unresolved
motion of emitting ions: either fluid motions (unresolved flows or waves) or
motion of accelerated non-thermal ions \citep[e.g.][]{2014ApJ...787...86J}. 
Large-scale fluid motions lead to
resonance broadening \citep[e.g.,][]{2012ApJ...754..103B,2012PhRvE..86f6308W}.
Considering first random motions in the transverse ($\perp_2$) direction, we
suggest that the small scale density fluctuations \citep[mostly near the inner
scale $q_i^{-1}$ of density turbulence responsible for radio-wave
scattering;][]{2023ApJ...956..112K} are advected by large-scale random plasma
motions with speeds corresponding to the outer scale of the
turbulence. Within this framework, the velocity fluctuations $\langle
v_{\perp_2}^2\rangle$ have a line broadening effect that is identical to that of
non-thermal ion velocities, and so can be modeled by replacing the Dirac
delta-function resonance condition by a finite-width Gaussian characterized by a
turbulent velocity $v_{\perp}$:

\begin{equation}
  \delta\left(\Omega (\vec{q})- \vec{q} \cdot \vec{v}_g\right)
  \rightarrow \frac{1}{\sqrt{2 \pi \, q_{\perp_2}^2 \langle v_{\perp_2}^2\rangle}}
  \exp \left[-\frac{(\Omega(\vec{q}) - \vec{q} \cdot \vec{v}_g)^2}{2 \, q_{\perp_2}^2 \langle v_{\perp_2}^2\rangle}\right] \,\,\, ,
\end{equation}
where $\langle v_{\perp_2}^2\rangle$ is the variance of large-scale motion
velocities. This is an application of a random sweeping hypothesis
\citep{1975JFM....67..561T} to radio scattering measurements.

The presence of random motions superimposed on the large-scale flows thus gives
diffusion tensor components

\begin{equation}\label{eq:D_random_plus_flows_general}
  D_{i j}=\frac{\pi \omega_{p e}^4}{4 \, \omega^2} \int q_i \, q_j \, S(\vec{q}) \,
  \frac{1}{\sqrt{2 \pi \, q_{\perp_2}^2 \langle v_{\perp_2}^2\rangle}}
  \exp \left[-\frac{(\Omega (\vec{q}) - \vec{q} \cdot \vec{v}_g)^2}{2 \, q_{\perp_2}^2 \langle v_{\perp_2}^2\rangle}\right]\, \frac{d^3 q}{(2 \pi)^3} \,\,\, .
\end{equation}

Integrating in the approximation $q_{\perp_2}^2 \langle v_{\perp_2}^2\rangle \ll
q^2c^2$ and taking $\Omega =0$ we obtain (Equation~\eqref{eq:random-static}) the
diffusion tensor

\begin{equation}\label{eq:d_random}
\mmatrix{D}=\frac{\pi \omega_{p e}^4}{16 \, \omega^2 c} \, \overline{q \, \eps^2}\;
\left(\begin{array}{ccc}
\alpha^2  & 0 & 0 \\
0 & 1 & 0 \\
0 & 0 & \frac{\langle v_{\perp_2}^2 \rangle}{c^2}
\end{array}\right) \,     
\end{equation}
leading to a frequency broadening that is mathematically similar to
Equation~\eqref{eq:D_vperp}, but where $\langle v_{\perp_2}^2 \rangle$ now
represents random velocity fluctuations. 

Similar considerations apply to random motions in the parallel (i.e., radial)
direction, with a factor of $\alpha^2$ applied, so that if there are random
motions in both directions,

\begin{equation}\label{eq:d_random_par_perp}
\mmatrix{D}=\frac{\pi \omega_{p e}^4}{16 \, \omega^2 c} \, \overline{q \, \eps^2}\;
\left(\begin{array}{ccc}
\alpha^2  & 0 & 0 \\
0 & 1 & 0 \\
0 & 0 & \frac{\alpha^2 \, \langle v_\parallel^2 \rangle \, + \, \langle v_{\perp_2}^2 \rangle}{c^2}
\end{array}\right) \,\,\, .     
\end{equation}
Again, since $\alpha <1$, perpendicular motions are more effective at frequency
broadening than radial (parallel) motions.

\section{Observed frequency broadening}\label{sec:observations}

\begin{figure}[h]
  \centering
  \includegraphics[width=0.8\textwidth]{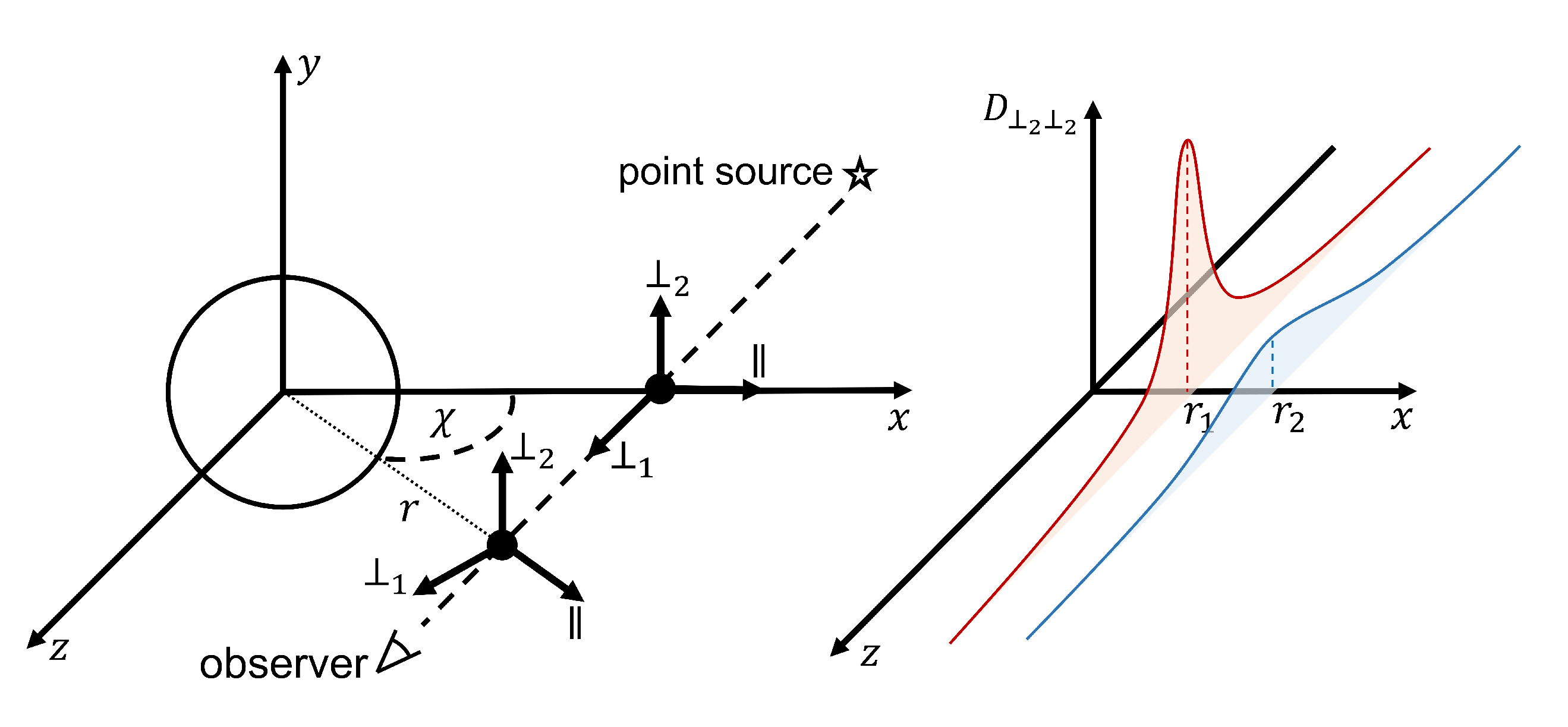}
  \caption{The left panel shows the Sun-centered coordinate system and
  its relation to heliocentric distance and the line of sight from a distant
  point source. The broadening of point sources is calculated as an integral
  along $z$. The right-hand panel qualitatively shows how the $D_{\perp_2\perp_2}$ diffusion
  tensor component (see e.g., Equation~\eqref{eq:D_ellastic}) varies along $z$,
  as illustrated for sources at two different heliocentric distances $r(z=0)$,
  with $r_1 <r_2$.}
  \label{fig:point_source_size}
\end{figure}

To compare with the observations, 
we note that radio wave with wavevector $\mathbf{k}$ is propagating along the
$z$-direction. We assign the $\parallel$ direction with the (assumed radial)
solar magnetic field $\vec{B}$ (Figure~\ref{fig:point_source_size}). We also
assign the $\perp_1$ direction to the perpendicular direction that is aligned
with the wave propagation direction at $z=0$, and $\perp_2$ to the perpendicular
direction that is orthogonal to both $\parallel$ and $\perp_1$, i.e.,
perpendicular to the projection of the radial direction on the plane of the sky.
The right-handed $(\parallel, \perp_2, \perp_1)$ coordinate system is obtained
by rotating the $(x,y,z)$ coordinate system by an angle $(- \chi)$ around the
$y$-axis (Figure~\ref{fig:point_source_size}).

Analogously to the results from the three previous subsections, the variance of
the wavenumber $k$ along the path of the radio wave due to motions in the plane
of the sky (here denoted by $v_{\perp \vec{k}}$) in the solar atmosphere can be
written as

\begin{equation}\label{eq:dk_z2}
\frac{d\langle k_z^2\rangle}{dt}= 2 D_{zz} = 
\frac{\pi \omega_{p e}^4}{8 \, \omega^2 \, c} \, \overline{q \, \eps^2} \,\frac{ \alpha^2 \langle v_\parallel^2 \rangle \cos^2 \chi  +\langle v_{\perp_2}^2 \rangle + \langle v_{\perp_1}^2 \rangle \sin^2 \chi}{c^2} \,\,\, ,
\end{equation}
where $\chi(z)$ is the angle between the radial direction of the magnetic field
and the $x$-axis (see Figure ~\ref{fig:point_source_size}). For perpendicular
motions that are dominated by gyrotropic turbulence,  $\langle v_{\perp_1}^2
\rangle = \langle v_{\perp_2}^2 \rangle = \langle v_\perp^2 \rangle$, this can
be written as


\begin{equation}\label{eq:dk_z2_2}
\frac{d\langle k_z^2\rangle}{dt}=\frac{\pi \omega_{p e}^4}{8 \, \omega^2 \, c} \, \overline{q \, \eps^2} \,
\frac{ v^2_{\perp \vec{k}}}{c^2} 
\,\,\, ,
\end{equation}
where

\begin{equation}\label{eq:v2perpk}
  v^2_{\perp \vec{k}} = \sqrt{ \alpha^2 \, \langle v_\parallel^2 \, \rangle \, \cos^2 \chi  + \langle v_{\perp}^2 \rangle  \, (1 + \sin^2 \chi) }  
\end{equation}
represents the weighted sum of all motions
perpendicular to $\mathbf{k}$, i.e., in the $(x,y)$ plane of the sky, and the
$\langle  v^2_\parallel \rangle$ term is the sum in quadrature of both steady
flows and random velocities in the parallel direction. 
Both random
motions and oscillations with the same phase speed contribute at the same
level.
Such motions and/or oscillations of density fluctuations in the plane of
the sky are perpendicular to the direction of radio wave propagation $\vec{k}$;
they hence lead to a change in wavenumber $\Delta \vec{k}$ that is aligned with
$\vec{k}$ and so to a change in the magnitude $|\vec{k}|$, i.e., to frequency
broadening. In the limit $\omega \gg \omega_{pe}$, the group velocity of the
radio wave $\vec{v}_{\rm gr} = \partial \omega/\partial {\vec k} = c^2 \,
\vec{k}/\omega \simeq c$. The frequency broadening rates per unit travel
distance $v_{\rm gr} \, dt$ along the direction of propagation $z$ can be
written \citep[similarly to][]{2023ApJ...956..112K} as

\begin{equation}\label{eq:int_los_df}
    \frac{\left\langle\Delta f^2\right\rangle}{f^2} \, = 
    \int_{los}\frac{1}{k_z^2} \, \frac{d\left\langle k_z^2\right\rangle}
    {c \, dt} dz = 
    \int _{los}\frac{\pi}{8} \, \frac{v_{\perp \mathbf{k}}^2}{c^2}  \, \frac{\omega_{p e}^4}{\omega^4} \, \overline{q \, \epsilon^2}(r) \, dz \,\,\, ,
\end{equation}
which can be integrated for known $v_{\perp \mathbf{k}}^2$. The right panel 
of Figure~\ref{fig:freq_broadening_obs} shows the predicted 
\citep[taking $n^2 \, \overline{q \, \eps ^2}$ from ][]{2023ApJ...956..112K} 
broadening for
a typical perpendicular speed $v_\perp=30$~km~s$^{-1}$ from non-thermal line measurements, 
and $v_\parallel=\sqrt{v_s^2 + v_{\mathrm{sw}}^2}$ where the sound speed $v_s$ is given
by Equation~\eqref{eq:sound_speed} and the solar wind speed $v_{\mathrm{sw}}$ is given by Equation~\eqref{eq:v_sw}. 
Importantly, the result does not depend on density model,
but on the strength of density fluctuations $n^2 \, \overline{q \, \eps ^2}$ and
the plasma velocities.

Equation \eqref{eq:int_los_df} can be also written approximately as  in Appendix~\ref{sec:analytic_df} or noting that the largest contribution to frequency broadening comes from the high density region near $z=0$ (Figure~\ref{fig:point_source_size}), and hence, to a good approximation, 
we can take $\chi \simeq 0$ in equation~\eqref{eq:int_los_df}.  
Thus, we can write
$v^2_{\perp \vec{k}}\simeq \alpha^2 \langle v_\parallel^2 \rangle + \langle
v_{\perp}^2 \rangle$ taking values at $z=0$. The frequency broadening integrated
over the path of the radio wave is now given by

\begin{equation} \label{eq:broadenning_approx}
  \frac{\langle \Delta f ^2\rangle}{ f^2 } \simeq 
  \frac{\pi}{8} \,  \frac{v_{\perp \mathbf{k}}^2}{c^2 \, \omega^4} \,\int _{los} \,
 \omega _{pe}^4\, \overline{q \, \eps ^2} \, dz = \frac{2 \pi^3 e^4}{m_e^2 \, c^2 \, \omega^4}\,  v_{\perp \mathbf{k}}^2 \, \int_{los} \,
 n^2 \, \overline{q \, \eps ^2} \, dz \,\,\, ;
\end{equation}
i.e.,

\begin{equation} \label{eq:broadening_final}
  \frac{\Delta f}{ f } \simeq 
 \frac{1}{(8 \pi)^{1/2}} \, \left ( \frac{e^2}{m_e \, c} \right ) \,  \left ( \int _{los} \,
 n^2 \, \overline{q \, \eps ^2} \, dz \right )^{1/2} \, \frac{v_{\perp \mathbf{k}}}{f^2} \,\,\, .
\end{equation}
which shows that the fractional frequency broadening $\Delta f/f$ 
depends on the carrier frequency $f$ as $1/f^2$ and is
determined by motions in the plane of the sky. 
Although both parallel and
perpendicular velocities may be present, the parallel velocities (both steady
flows and random motions) are weighted by the anisotropy parameter $\alpha <1$
\citep[cf. the expression for frequency broadening in an isotropic plasma;
Equation~(36) of][]{2019ApJ...873...33B}.
Knowing the anisotropy factor $\alpha$ and the $n^2 \, \overline{q \, \eps ^2}
(z)$ density fluctuation profile from independent measurements, one can deduce
the characteristic speeds of density fluctuations using
Equation~\eqref{eq:broadening_final}. Indeed, using an analytic approximation
for $n^ 2 \, \overline{q \, \eps ^2}$ derived from solar observations
\citep{2023ApJ...956..112K}, one can find an approximate analytical expression
for $\Delta f$, which is presented in Appendix~\ref{sec:analytic_df}. 

\section{Frequency broadening measurements}\label{sec:freq_broad_measurements}

\begin{figure}
\centering
\includegraphics[width=0.49\textwidth]{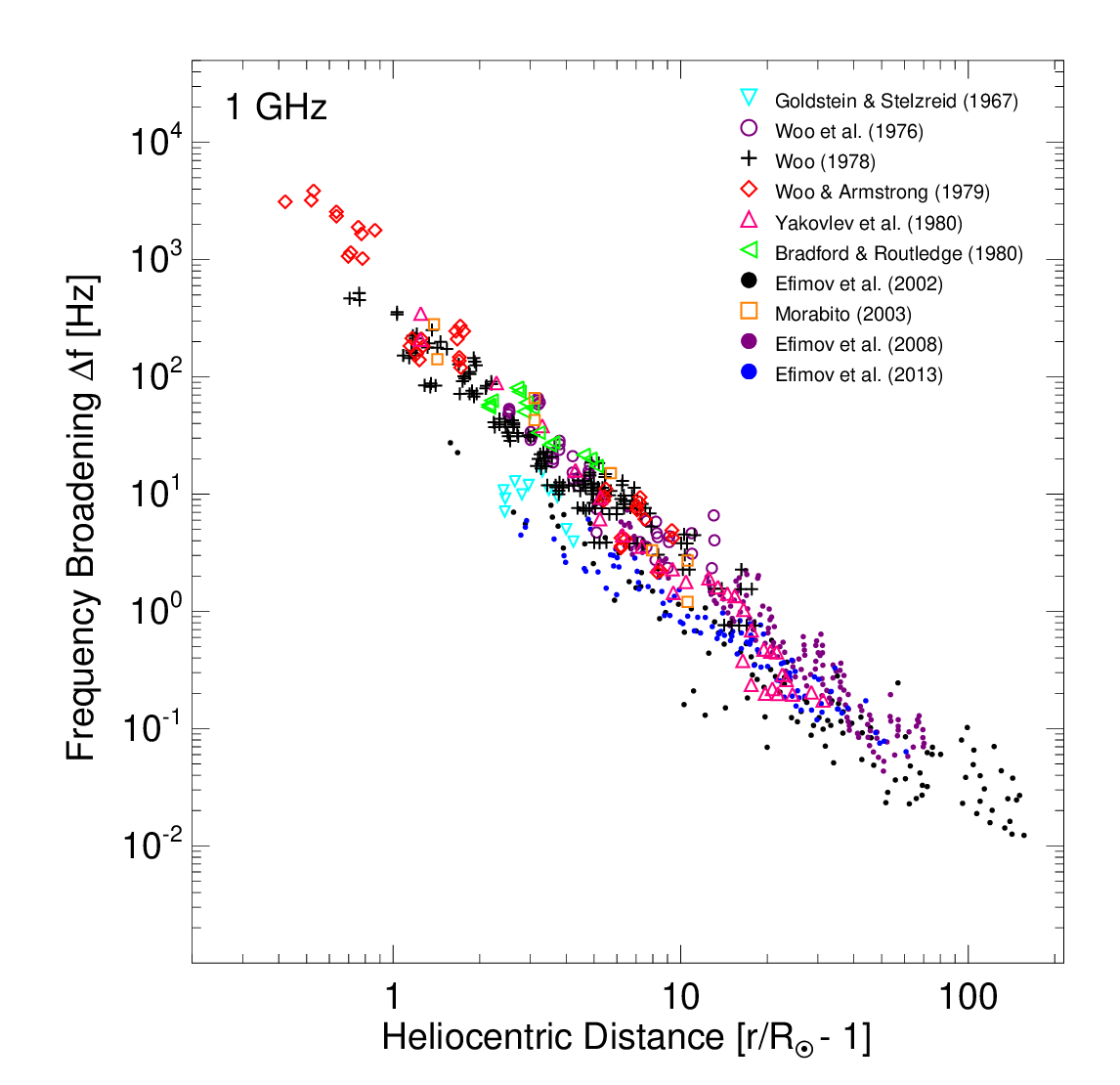}
\includegraphics[width=0.49\textwidth]{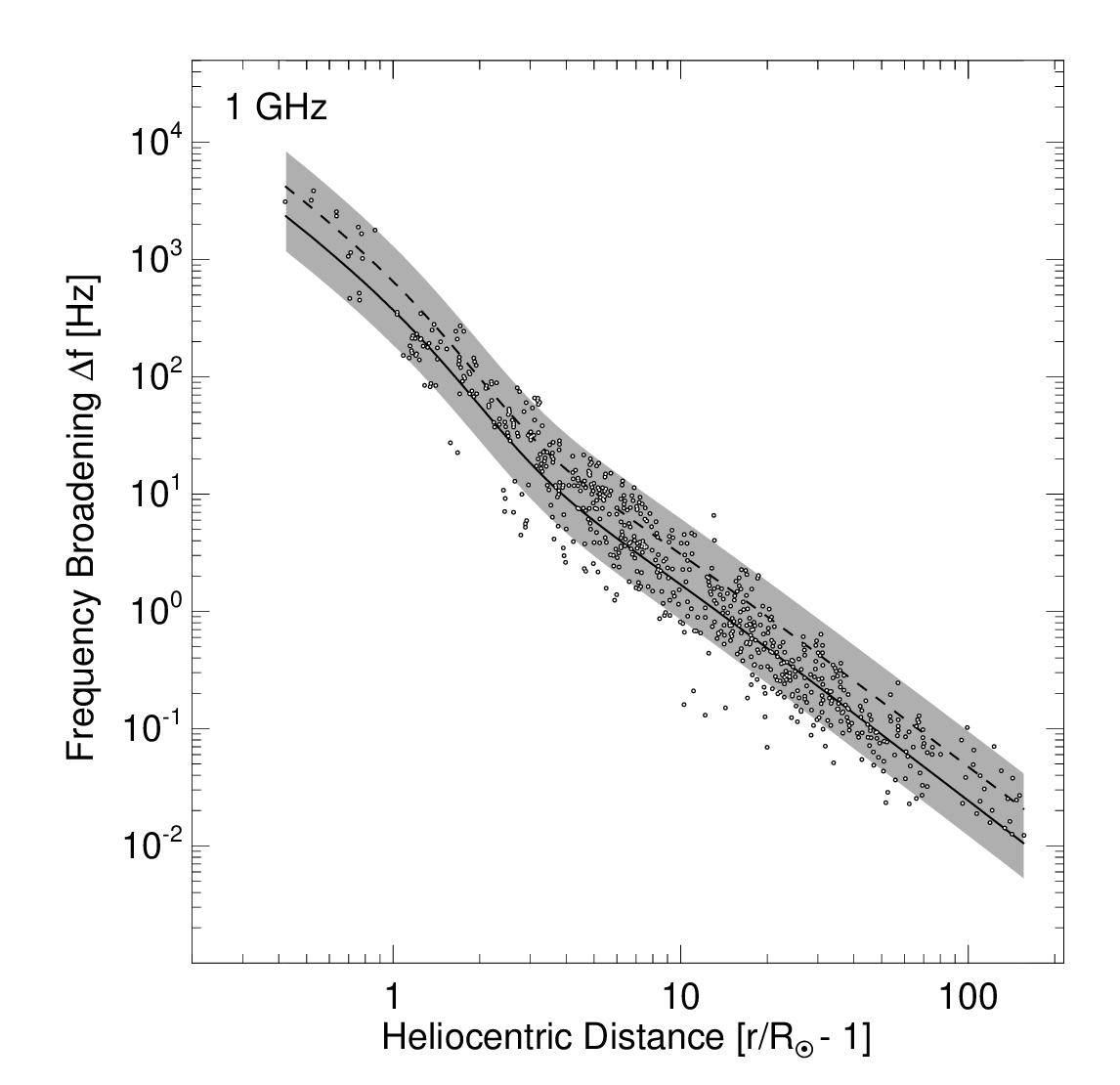}
   \caption{{\textbf{ Left:}} Observed spectral broadening $\Delta f =
  \sqrt{\langle\Delta f^2\rangle}$ (the square root of the variance) of
  spacecraft signals observed through the corona from various studies, where
  each carrier signal is scaled to $f=1$~GHz using $\Delta f/f \propto 1/f^2$.
  The conversions applied to the different measures to retrieve the standard
  deviation are noted in Appendix~\ref{sec:conversion-to-sigma}.
  {\textbf{Right:}} Form of $\Delta{f}$ derived from
  Equations~\eqref{eq:v2perpk} and~\eqref{eq:int_los_df}, for $v_{\perp}=30$~km s$^{-1}$ (from the non-thermal line broadening measurements in Appendix \ref{sec:non-thermal_vel}), and $v_\parallel=\sqrt{v_s^2 + v_{\mathrm{sw}}^2}$, where the sound speed $v_s$ is given
  by Equation~\eqref{eq:sound_speed} and the solar wind speed $v_{\mathrm{sw}}$ is given by Equation~\eqref{eq:v_sw}. The
  solid and dashed lines show $\Delta{f}$ derived using $1\times\overline{q \,
  \epsilon^2}$ for $\alpha=0.25$ and $\alpha=0.4$, respectively, while the grey
  area shows the range in $\Delta{f}$. The lower bound is given by $1/2 \times
  \overline{q \, \epsilon^2}$ for $\alpha=0.25$, and the upper bound is given by
  $2 \times \overline{q \, \epsilon^2}$ for $\alpha=0.4$.}
  \label{fig:freq_broadening_obs}
\end{figure}

Frequency broadening observations (for details see
Appendix~\ref{sec:conversion-to-sigma}) have been conducted a number of times
using signals from different spacecraft \citep{goldstein1967superior,
1976ApJ...210..593W, 1978ApJ...219..727W, 1979JGR....84.7288W,
1980MNRAS.190P..73B,
1980SvA....24..454Y,2002AdSpR..30..453E,2003ITAP...51..201M,2008AdSpR..42..117E,
2013CosRe..51...13E}. The left panel of Figure~\ref{fig:freq_broadening_obs}
shows the compilation of $1\sigma$ frequency broadening (the square root of the
variance, $\Delta f \equiv \sqrt{\langle\Delta f^2 \rangle}$\,) of spacecraft
signals as a function of heliocentric distance. For observation at different
frequencies, the broadened quantity is scaled to 1~GHz using
$\Delta{f}_{1\mathrm{GHz}}=\Delta{f}_{\mathrm{obs}}\left(f_{\mathrm{obs}}
[\mathrm{GHz}]\right)^2$. Appendix~\ref{sec:conversion-to-sigma} provides
information on the conversion of other reported broadening measures. The trend
of $\Delta f$ with heliocentric distance follows a broken power-law, with a
 steeper power law index $\sim -2$ below $\sim 3 \, R_\odot$,
transitioning to a somewhat flatter power-law index of approximately $-1.7$
above $\sim$$10$~R$_\odot$. 

\begin{figure}
\centering
   \includegraphics[width=0.33\linewidth]{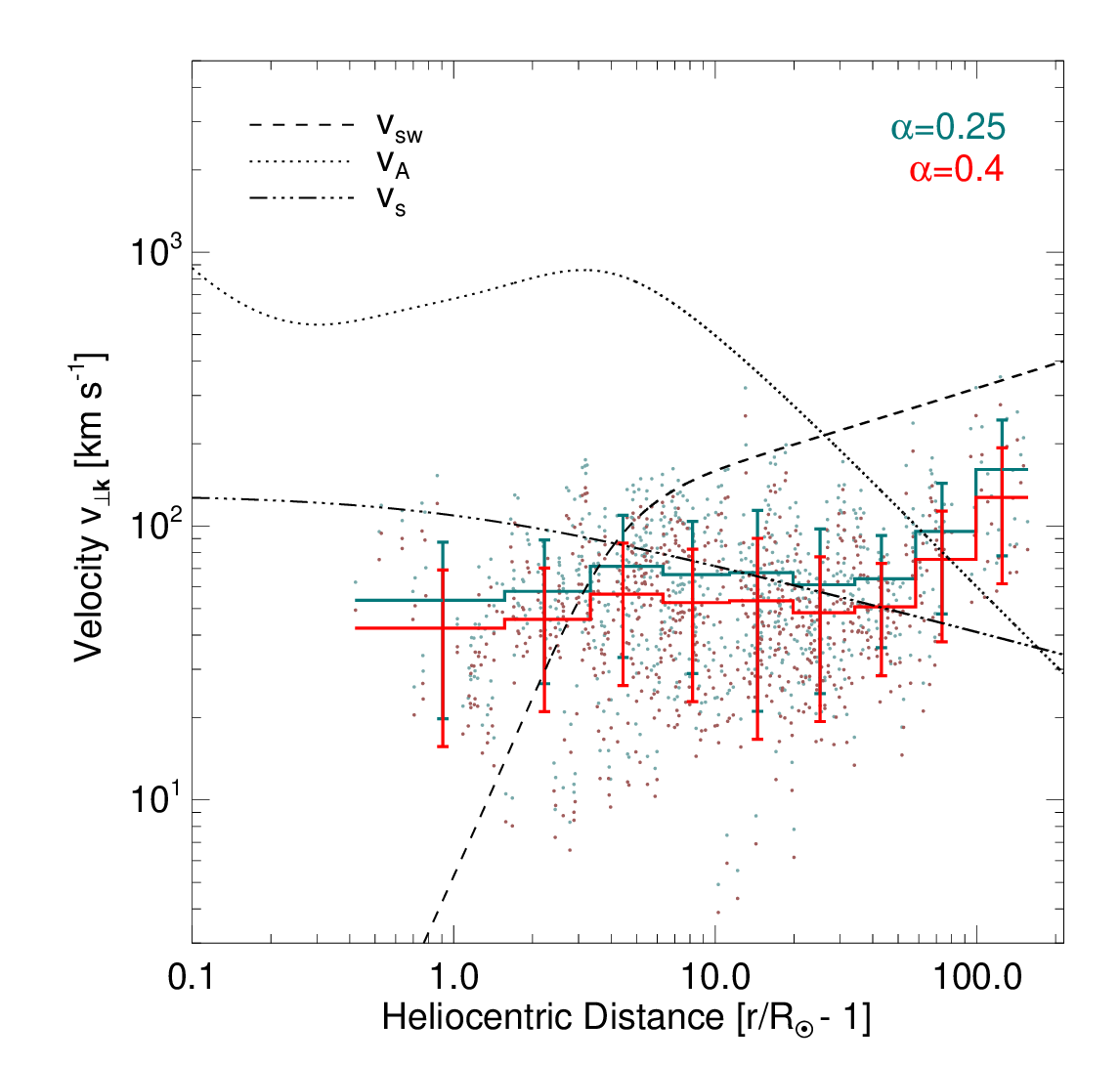}
   \includegraphics[width=0.33\linewidth]{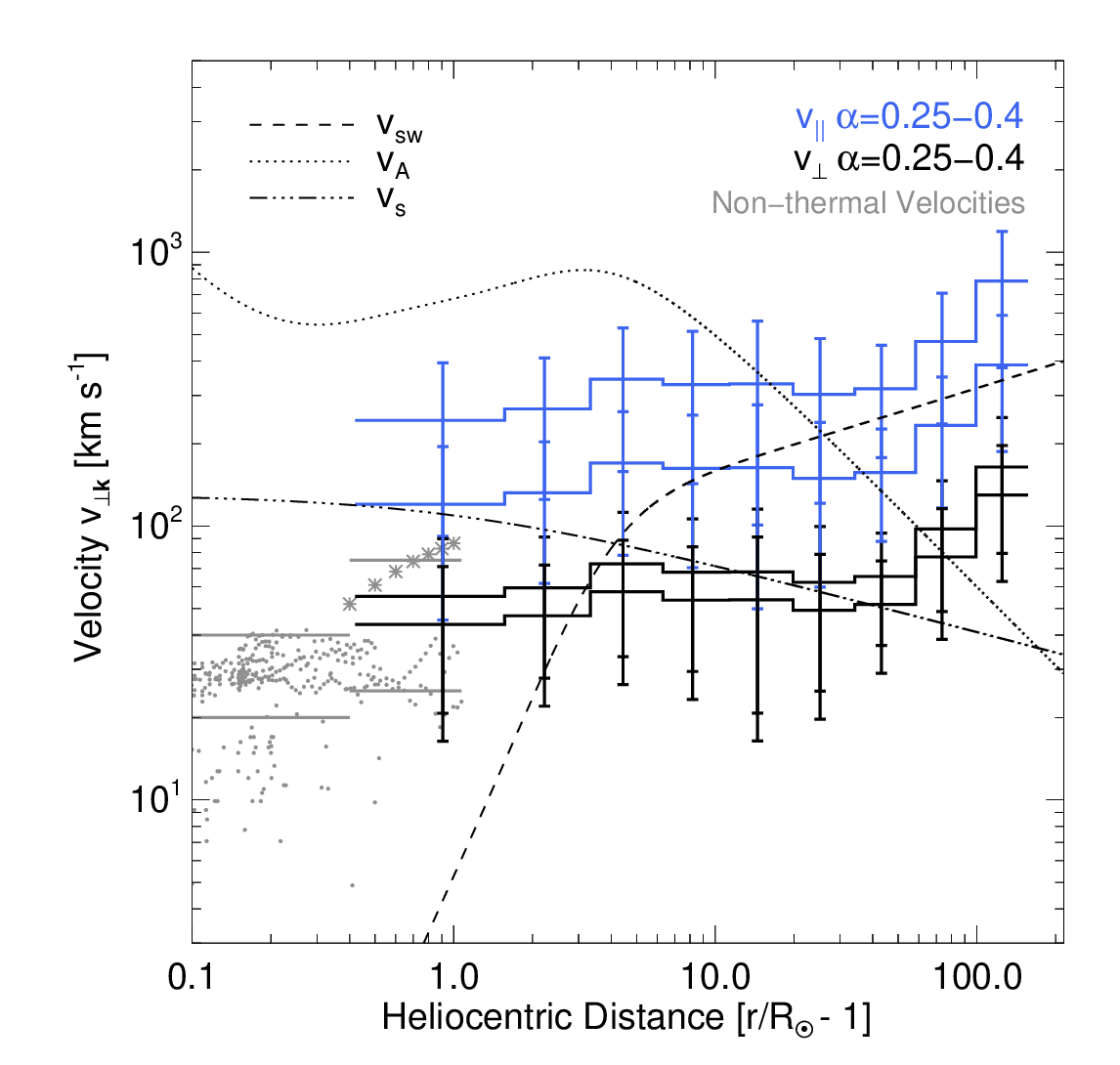}
   \includegraphics[width=0.33\linewidth]{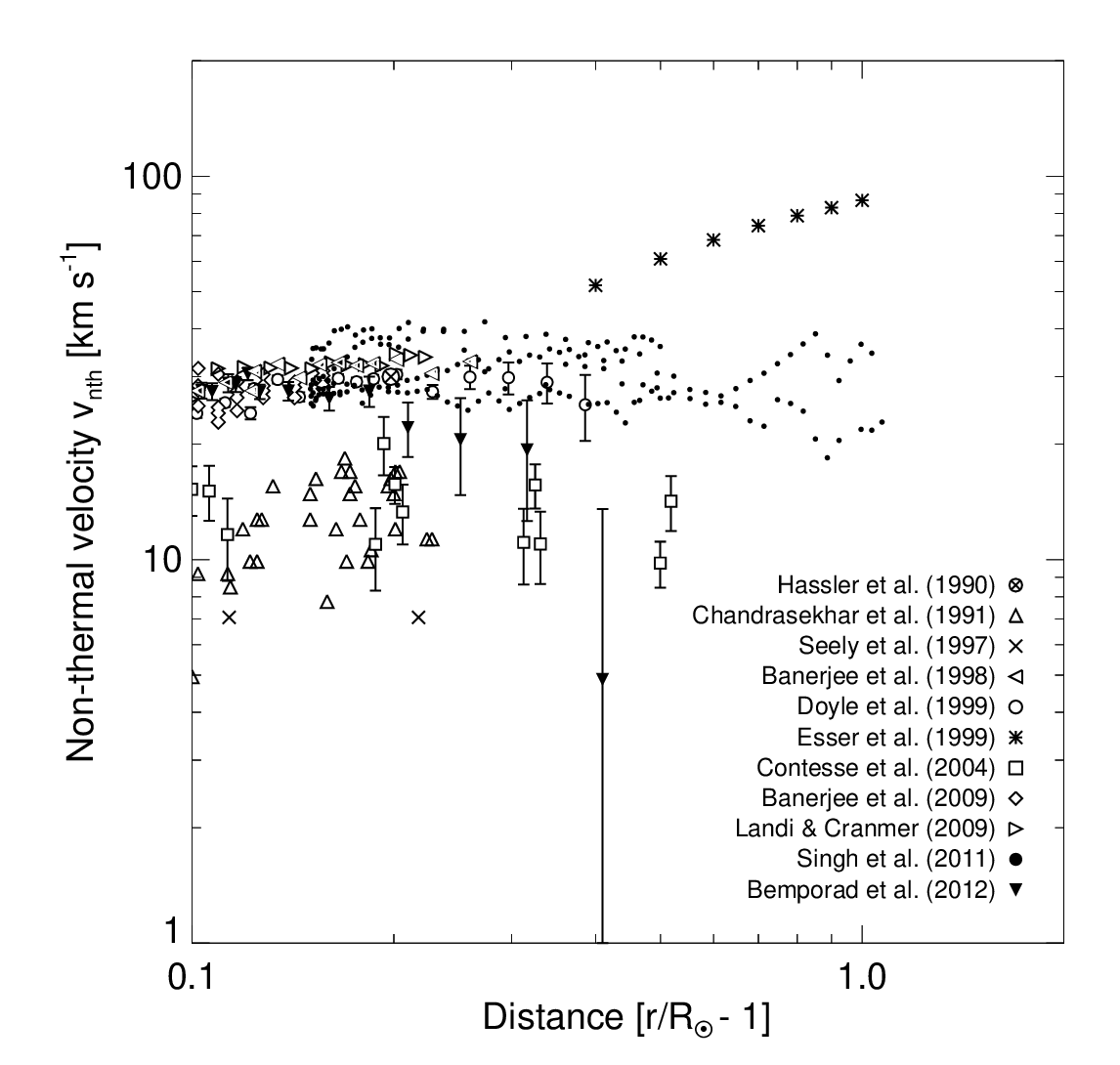}
\caption{\textbf{Left:}
Plane-of-sky velocity $v_{\perp \vec{k}}$ calculated using frequency broadening
measurements from Figure~\ref{fig:freq_broadening_obs},
Equation~\eqref{eq:broaden_vel}, and the $n^2 \, \overline{q \, \eps ^2}$ values
at various distances $r$ derived from measurements of other phenomena, such as
angular broadening of extra-solar sources and the location, size and timing of
solar radio bursts \citep{2023ApJ...956..112K}. The green and red points show
the conversion of individual $\Delta{f}$ data points from
Figure~\ref{fig:freq_broadening_obs}, for different values of $\alpha$, with
binned averages and weighted uncertainties on each bin. \textbf{Middle:}
Parallel and perpendicular velocities $v_{\parallel}$ (blue) and $v_{\perp}$
(black) required to solely explain the frequency broadening measurements in
Figure~\ref{fig:freq_broadening_obs}. The grey dots and stars show a summary of
the measured values of the non-thermal velocity standard deviation from the
right panel. Also shown is the solar wind speed $v_{\rm SW}$
(Equation~\eqref{eq:v_sw}), the ion-sound speed $v_s$
(Equation~\eqref{eq:sound_speed}), and the Alfv\'en speed from
Equation~\eqref{eq:alfven_speed}, obtained using the density and magnetic field
models in Equations~(A1) and~(A2) of \cite{2023ApJ...956..112K}. \textbf{Right:}
$1\sigma$ non-thermal velocities $v_{\mathrm{nth}}$ from line-of-sight Doppler
broadening of coronal lines (see Appendix~\ref{sec:non-thermal_vel} for
details).}
\label{fig:df2velEff}
\end{figure}

Instead of assuming a flow (or turbulent) velocity value, we can alternatively
use the measured frequency broadenings to determine the associated velocity, by
rewriting Equation~\eqref{eq:broadening_final} in the form
\begin{equation} \label{eq:broaden_vel}
 v_{\perp\vec{k}} \equiv \sqrt{\alpha^2 \, \langle v_\parallel^2 \rangle  \, +\langle  v_{\perp}^2 \rangle  } 
 \simeq  \frac{( 8 \pi )^{1/2}}{c \, r_o} \, \frac{f^2 }{\left ( \int_{los} n^2 \, \overline{q \, \eps ^2}  \, dz \right )^{1/2}} \,\, \frac{\Delta f}{f} \,\,\, ,
\end{equation}
where $r_o = e^2/m_e c^2$ is the classical electron radius and is 
evaluated taking $n(r[z])\overline{q \, \epsilon^2} \, (r[z])$ from
\citet{2023ApJ...956..112K}. 
In the left panel of Figure~\ref{fig:df2velEff} we show $v_{\perp \vec{k}}$ (in
km~s$^{-1}$) as a function of heliocentric distance $r$, for two different
values of the anisotropy parameter $\alpha$. Further, by taking $v_{\perp} = 0$
or $v_\parallel = 0$, one can obtain an upper limit on the magnitude of the
remaining component of $v_{\perp \mathbf{k}}$. The middle panel of
Figure~\ref{fig:df2velEff} compares these maximum values of $v_\parallel$ and
$v_{\perp}$ with various other speeds, including the solar wind speed $v_{\rm
SW}$, the sound speed $v_s$, the Alfv\'en speed $v_A$, and nonthermal velocities
deduced from UV spectral line broadening observations. 
These reference speeds are calculated as follows:

\begin{itemize}

  \item {\it Sound speed:} The electron temperature of the solar wind is
  observed to decrease with heliocentric distance: $T_e \propto r^{-(0.3 \, - \,
  0.7)}$ \citep[e.g.,][]{2015JGRA..120.8177A}. If we model the temperature as
  $T_e \simeq~2~\times~10^6 \, (r/R_\odot -1)^{-0.5}$~K, then the sound speed
  $v_s\simeq \sqrt{k_BT_e/m_i}$ varies with heliocentric distance $r$ as

\begin{equation} \label{eq:sound_speed}
    v_s (r) \simeq {130} \, \left  ( \frac{r}{R_\odot} -1 \right )^{-0.25} \,\, \text{km s$^{-1}$} \,\,\, .
\end{equation}

  \item {\it Solar wind speed:} In the spherically symmetric expanding corona
  \citep{1958ApJ...128..664P}, mass conservation $v_{sw} \, r^2 \, n(r) =
  \mbox{const}$ requires that, with a typical solar wind speed of
  $400$~km~s$^{-1}$ at 1~au,

\begin{equation}\label{eq:v_sw}
   v_{\text{SW}} (r) \approx 400 \, \left(\frac{n \, (\mbox{1~au})}{n(r)}\right) \, \left(\frac{\mbox{1~au}}{r} \right)^2 \, \text{km s$^{-1}$} \,\,\,  ,
\end{equation}
where $n(r)$ is the plasma density.

\item {Alfv\'en speed:} The Alfv\'en speed 

\begin{equation}\label{eq:alfven_speed}
    v_A (r) = \frac{B(r)}{\sqrt{4 \pi m_i \, n(r)}}
\end{equation}

is obtained using the magnetic field and density models in Equations~(A1)
and~(A2) of \cite{2023ApJ...956..112K}.

\item {\it Nonthermal velocities:} Nonthermal turbulent velocities are inferred
through measurement of the excess width of EUV coronal spectral lines compared
to their thermal widths, and are often interpreted as evidence of perpendicular
velocity fluctuations at speeds of a few tenths of the Alfv\'en speed
\citep[e.g.,][]{1998SoPh..181...91D,2011SoPh..270..213S}. The inferred turbulent
velocities are dependent on the assumed ion temperature
\citep{1997ApJ...484L..87S}. Measurement of the width of spectral lines in the
radio domain also provide a (temperature-independent) measure of velocity
fluctuations. Figure~\ref{fig:df2velEff} also shows various measurements of
$1\sigma$ nonthermal velocities at different heliocentric distances; this
information is also summarized in the middle panel of
Figure~\ref{fig:df2velEff}.

\end{itemize}

The middle panel of Figure~\ref{fig:df2velEff} shows clearly that the velocities
deduced from frequency broadening measurements become dominated by the solar
wind speed at large heliocentric distances $r>10 \, R_\odot$. However, closer to
the Sun at $r \lapprox 10 \, R_\odot$, the solar wind speed contribution is much
smaller than the inferred $v_{\perp \mathbf{k}}$ speeds, whether radial
velocities $v_\parallel$ in the range (100-300)~km~s$^{-1}$, or perpendicular
motions $v_{\perp}$ in the range (25-75)~km~s$^{-1}$, or a combination of such
motions are considered. Both these inferred speed ranges are well below the
Alfv\'en speed; however, perpendicular motions of this magnitude are quite
consistent with the nonthermal speeds deduced from the observed widths of
UV spectral lines. The next section discusses such nonthermal turbulent motions 
and their possible role in coronal heating.

\section{Velocity fluctuations and Alfv\'en wave
cascade}\label{sec:alfven_cascade}

Velocity fluctuations (non-thermal velocities of emitting ions) 
along of line of sight are often interpreted as manifestations
of perpendicular to magnetic field velocities
\citep[e.g.,][]{1998SoPh..181...91D} 
and (as we have seen above) frequency broadening of radio signals. 
Such motions are commonly interpreted as Alfv\'en
waves, which can undergo turbulent cascade to smaller scales
\citep[e.g.,][]{1978RvGSP..16..689H,1982SSRv...33..161L,1995ARA&A..33..283G,1995SSRv...73....1T}.
The power per unit mass (erg~g$^{-1}$~s$^{-1}$) available to be deposited
through such a Kolmogorov cascade in strong MHD turbulence
\citep{1995ApJ...438..763G} is estimated to be

\begin{equation}\label{eq:dE_dt_vperp}
\epsilon_{\ell_\perp} \simeq \frac{ v_{\perp}^2 }{\tau} \simeq  \frac{ v_{\perp}^3}{\ell_\perp} \,\,\, ,
\end{equation}
where the characteristic cascade time is $\tau = \ell_\perp/v_{\perp}$, with
$\ell_\perp$ being a measure of the transverse correlation length
\citep{1986JGR....91.4111H}. Although $\ell_\perp$ is not measurable directly in
the corona, one can assume it to be comparable to the transverse size of a flux
tube \citep[see Equation~(4) in][]{1986JGR....91.4111H}:

\begin{equation}
    \ell_\perp =\frac{7.5 \times 10^8}{\sqrt{B}}\;\text{cm} \,\,\, ,
\end{equation}
where $B$ is in Gauss. This is also similar to the estimate
used in MHD simulations \citep[see Equation~(51) in][and subsequent discussion
therein]{2005ApJS..156..265C}. 

The quantity $\epsilon_{\ell_\perp}$ is the power per unit mass at the outer
scale of the inertial range; it is the rate at which energy enters the turbulent cascade
process at the largest scales, and it is a scale-invariant quantity within the
inertial range of turbulence. The value of $\epsilon_{\ell_\perp}$ often serves
as a measure of coronal heating via Alfv\'en turbulent cascade, 
or the specific energy rate associated with anisotropic MHD turbulence
\citep{1995ApJ...438..763G}. This energy input rate plays a role that is
conceptually similar to the solar flare scenario, in which an Alfv\'en
turbulence cascade (estimated from measured non-thermal velocities) is believed
to power particle acceleration in solar flares
\citep{2017PhRvL.118o5101K,2021ApJ...923...40S}.

\begin{figure}
\centering
\includegraphics[width=0.49\textwidth]{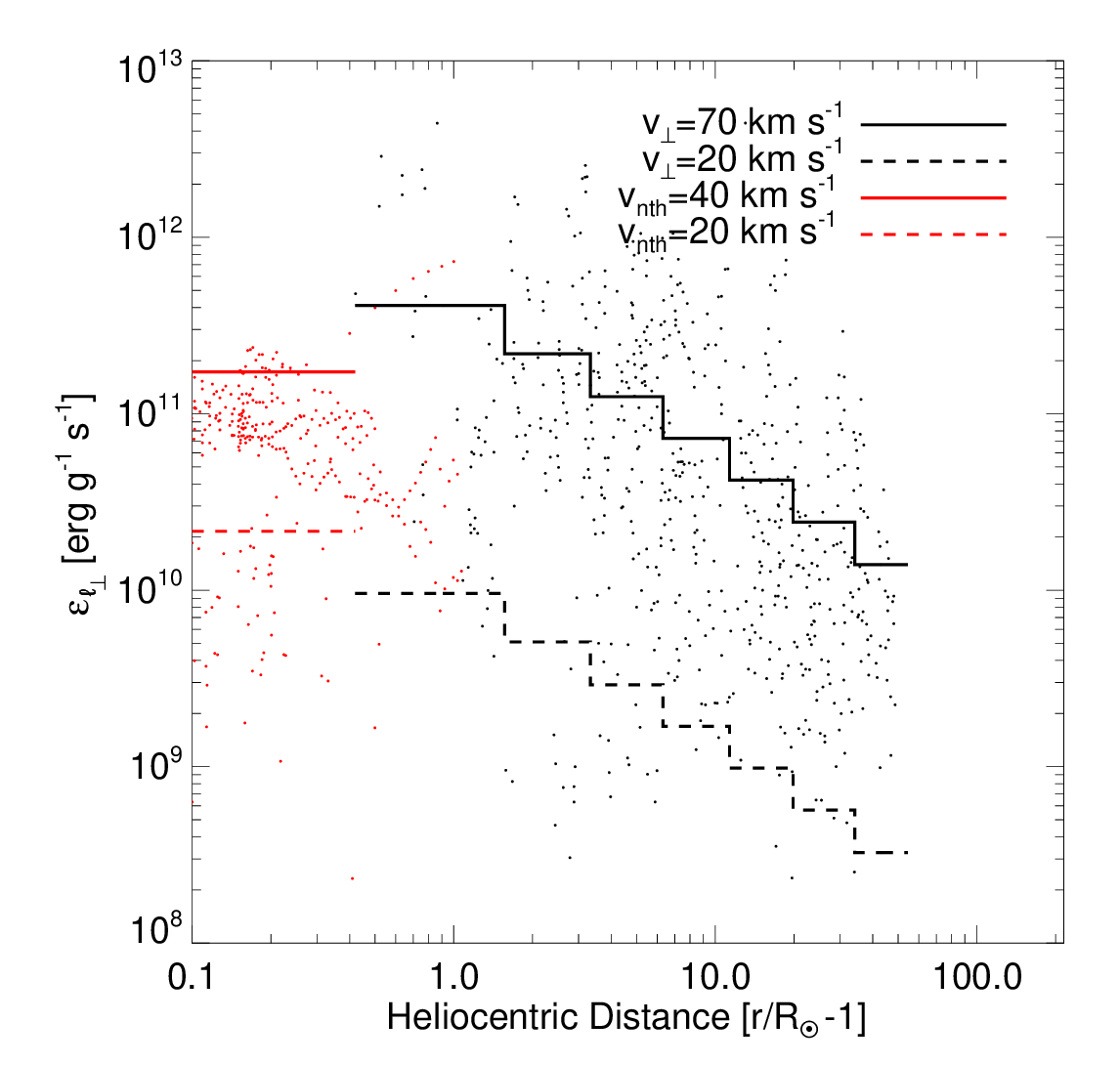}
\includegraphics[width=0.49\textwidth]{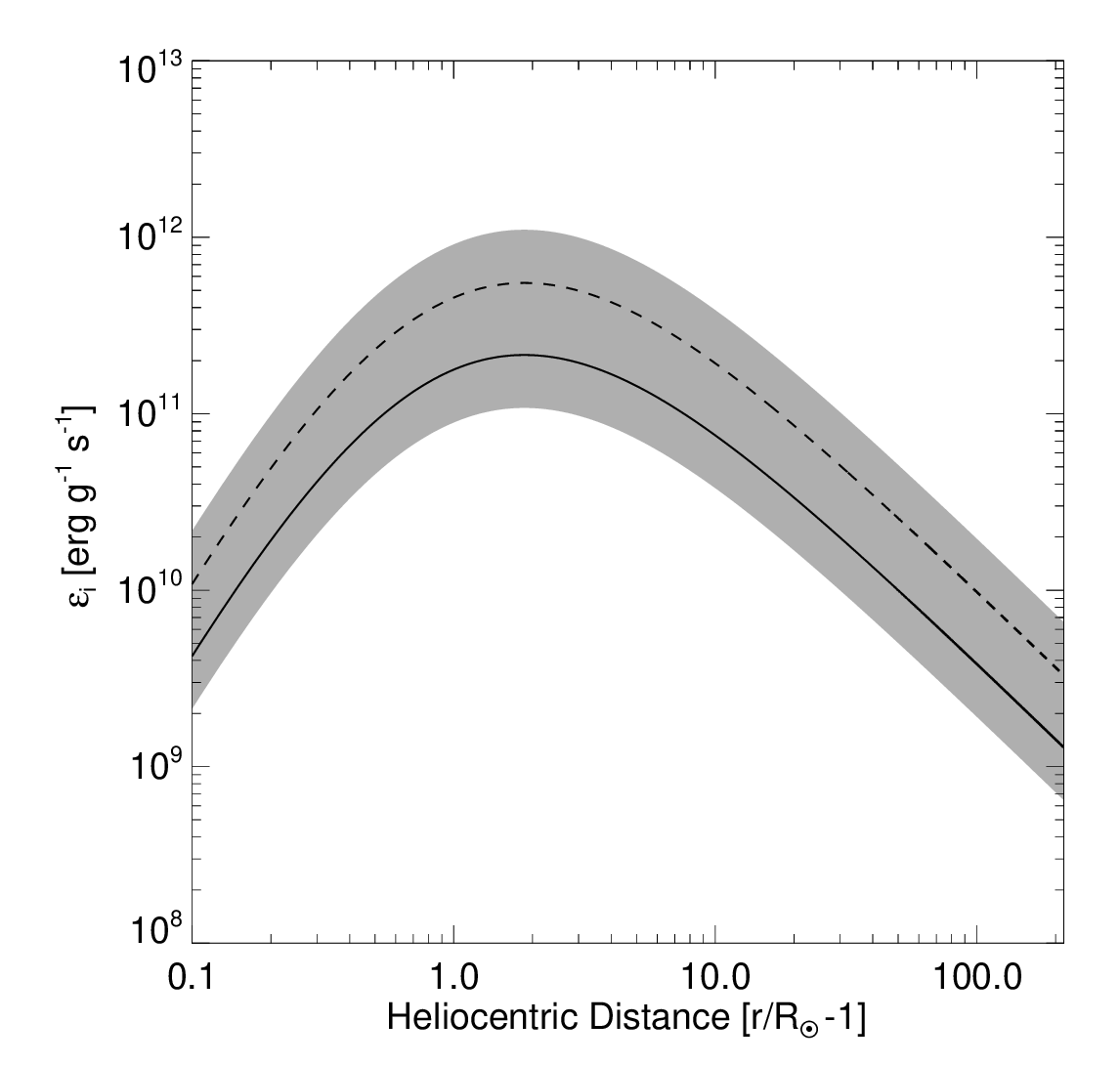}
  \caption{\textbf{Left:} Available power per unit mass, $\epsilon_{\ell_\perp}$
  (erg~g$^{-1}$~s$^{-1}$) from Equation~\eqref{eq:dE_dt_vperp}, using
  perpendicular velocity fluctuations from Figure~\ref{fig:df2velEff}. The solid
  and dashed lines correspond to the range (20-70)~km~s$^{-1}$ (black) from
  frequency broadening measurements, and (20-40)~km~s$^{-1}$ for $r < 1.4 \,
  R_\odot$ for non-thermal velocities from coronal lines (red). \textbf{Right:}
  Coronal heating rate per unit mass (erg~g$^{-1}$~s$^{-1}$) from Landau damping
  of ion-sound waves given by Equation~\eqref{eq:dE_dt2}. As in the right panel
  of Figure~\ref{fig:freq_broadening_obs}, the solid and dashed lines show $1
  \times\overline{q \, \epsilon^2}$ for $\alpha=0.25$ and $\alpha=0.4$,
  respectively, and the grey area corresponds to the range of values
  $[1/2,2]\times\overline{q\, \epsilon^2}$, considering both values of
  $\alpha$.}
  \label{fig:heating_rate}
\end{figure}

The left panel of Figure~\ref{fig:heating_rate} shows the values of
$\epsilon_{\ell_\perp}$ deduced from the perpendicular velocities inferred from
radio wave frequency broadening observations (Figure~\ref{fig:df2velEff}). They
suggest an energy cascade rate $\epsilon_{\ell_\perp} \simeq
10^{11}$~erg~g$^{-1}$~s$^{-1}$, a value that is similar to earlier estimates
\citep[e.g.,][]{1986JGR....91.4111H,2005ApJS..156..265C}.

\section{Damping of ion-sound waves}\label{sec:ion_sound_wave_damping}

Energy could be supplied to the corona and solar wind via absorption of the
energy contained in ion-sound or slow magneto-sonic waves \citep[e.g.,][and
references therein]{2020ApJ...891...51K}, that are often observed in the corona
\citep[e.g.,][]{1998ApJ...501L.217D,2009ApJ...696.1448W,2012A&A...546A..93G} and solar wind close to the Sun \citep{2024arXiv240314861Z}. 
For $q \, \lambda_{D e} \ll 1$ (where $\lambda_{De}$ is the Debye length), 
the spectral energy density of parallel propagating ion-sound waves $W_q$
(erg~cm$^{-3}$~[cm$^{-1}$]$^{-3}$]) is related to the spectrum of density
fluctuations $S(q)$ ([cm$^{-1}$]$^{-3}$]) by \citep[see, e.g., Appendix~C
in][]{2017SoPh..292..117L}

\begin{equation}
\frac{W_q^s}{n_e k_B T_e} \simeq \frac{\left|\delta n_e\right|_q^2 \, (q)}{n_e^2} \equiv S(q) \,\,\, .
\end{equation}
Parallel propagating ion-sound waves are strongly damped, especially in the
plasma with $T_i\simeq T_e$. The Landau damping rate $\gamma_q^s$ (s$^{-1}$) of
ion-sound waves with $v_s\simeq \sqrt{k_BT_e/m_i}$ 
is proportional to the wave frequency $\Omega_q^s = v_s \, q_\parallel$
\citep[e.g.,][]{1973ppp..book.....K,2012wop..book.....P}:

\begin{equation}\label{eq:landau_gamma}
\gamma_q^s = \sqrt{\frac{\pi}{8}} \, \Omega_q^s \, 
\left\{\sqrt{\frac{m_e}{m_i}}+\left(\frac{T_e}{T_i}\right)^{3/2} 
\exp \left[-\left(\frac{T_e}{2T_i}\right)\right]\right\} \,\,\, ,
\end{equation}
where $k_B$ is Boltzmann's constant and $T_e$ and $T_i$ are the electron and ion
temperatures, respectively. The first term on the right-hand-side of
Equation~\eqref{eq:landau_gamma} is the electron contribution, while the second
term is from protons. For $T_e \simeq T_i$ ion-sound waves are subject to very
strong damping, with a damping rate becoming

\begin{equation}\label{eq:ion-sound-damping-rate}
\gamma^s_q \simeq \sqrt{\frac{\pi}{8 \, e}} \, v_s \, q_\parallel \simeq 0.4 \, v_s \, q_\parallel
\end{equation}
that is a substantial fraction of the wave frequency. This strong damping of the
energy associated with ion-sound waves results in a volumetric
energy deposition rate (erg~cm$^{-3}$~s$^{-1}$)

\begin{equation}
    \frac{dE}{dt} = \int 2 \, \gamma^s_q  \, W^s_q \, d^3q  \,\,\, ,
\end{equation}
or, equivalently, a coronal heating rate per unit mass (erg~g$^{-1}$~s$^{-1}$)

\begin{equation} \label{eq:dE_dt2}
\epsilon_i=\frac{1}{\rho} \, \frac{dE}{dt} = \frac{2}{m_i n}\int \gamma^s_q \, W^s_q \, d^3q 
\simeq  0.8 \, v_s^3 \, \int |q_\parallel| \, S(\vec{q}) \, \frac{d^3q}{(2\pi)^3}
= 0.8 \, \alpha \, v_s^3 \, \overline{q \, \eps^2} \,\,\, ,
\end{equation}
where the third equality follows from the fact that the ion-sound waves
propagate along the (radial) magnetic field. The right panel of
Figure~\ref{fig:heating_rate} shows the heating rate given by
Equation~\eqref{eq:dE_dt2}. The heating profile has a shape that is similar to the heating functions often used in simulations \citep[e.g., Figure~7
in][]{2010ApJ...710..676C}, with a broad maximum at
(1-3)$ \, R_\odot$, consistent with the observed increasing temperature of the
solar corona out to this radius \citep{1997ApJ...482..510W}. 
The inferred energy deposition rate, integrated over the range of heights ($2-3R_\odot$) 
where it is most effective, 
corresponds to an energy flux, 
$\epsilon _i(2R_\odot) m_i n(2R_\odot) R_\odot \sim 6 \times 10^5$~erg~ cm$^{-2}$~s$^{-1}$ (600~W)
where $\epsilon _i(2R_\odot)\simeq 10^{11}$~erg~g$^{-1}$~s$^{-1}$ 
and $n(2R_\odot)\simeq 5\times 10^6$~cm$^{-3}$. 
We would note that while this value represents a useful constraint on the heating 
and dynamic energy terms in the corona, directly comparing its magnitude to such terms 
represents a considerable over-simplification of the modeling 
of coronal heating and/or solar wind acceleration. 
Nevertheless, we would point out that such an energy flux 
is broadly consistent with that required to balance energy losses 
and so heat the corona \citep[e.g.,][]{1977ARA&A..15..363W,1986JGR....91.4111H,1988ApJ...325..442W}.

The heating rate \eqref{eq:dE_dt2} is proportional to the quantity $\overline{q
\, \eps^2}$, which, as we have seen, can be inferred from observations related
to radio-wave scattering. As shown in Figure~1 of  \cite{2023ApJ...956..112K},
$\overline{q \, \eps^2}$ is dominated by fluctuations at short wavelengths near
the inner scale $q_i^{-1} \sim c/\omega_{pi}$, 
so that (their Equation~(29)):

\begin{equation}
    \overline{q \, \epsilon^2} \simeq 5 \, q_i \, \frac{\langle\delta n_i^2\rangle}{n^2} \,\,\, .
\end{equation}
The coronal heating rate per unit mass due to absorption at heliocentric
distance $r$ can therefore be expressed rather succinctly as $\epsilon_i (r)
\simeq 4 \, \alpha \,   q_i \, v_s^3 \, {\langle\delta n_i^2\rangle}/{n^2} $.

The quantities $\epsilon_{\ell_\perp}$ and $\epsilon_i$ are associated with very
different physical models, and they are associated with length scales that span
five orders of magnitude: the power generated in large-scale Alfv\'en motions
$\epsilon_{\ell_\perp}$ is dominated by scales $\ell_\perp (r=2 \, R_\odot)
\simeq 10^4$~km, while the energy dissipation rate $\epsilon_i$ due to ion-sound
wave damping is dominated by waves at the inner scale $q_i^{-1}$ of the
turbulence spectrum, which at $r = (2-3) \, R_\odot$ is of order $0.1$~km.
Despite this vast difference in characteristic scales, the quantities
$\epsilon_{\ell_\perp}$ and $\epsilon_i$ at $r\simeq (1-2) \, R_\odot$ are very
similar; indeed, they are identical within the error bars, with
$\epsilon_{\ell_\perp} \simeq \epsilon_i\simeq  10^{11}$~erg~g$^{-1}$~s$^{-1}$
(Figure~\ref{fig:heating_rate}). This result is both unexpected and tantalizing,
suggesting that the energy associated with large-scale magnetic field motions
can effectively cascade over the entire inertial range, eventually appearing as
small-scale ion-sound waves that are very effectively damped, causing plasma
heating. This intriguing result has significant implications for models of
coronal heating.

\section{Summary and Discussion}\label{sec:summary}

Using a density fluctuation model obtained from analysis of solar radio bursts,
combined with frequency broadening measurements from various spacecraft, we have
deduced the magnitude of the characteristic velocities in the solar corona and
the solar wind. The inferred velocities depend on the anisotropy of the density
turbulence. The amount of spacecraft signal broadening, and the anisotropic
density fluctuation inferred from solar burst data, tell a remarkably coherent
story about the level of density turbulence in the solar corona and the bulk
flow speeds present; the latter are consistent with previously published values
that employed different analysis techniques. The perpendicular velocities are
also consistent with the non-thermal speeds deduced from line-of-sight Doppler
broadening of spectral lines in the low corona. Interpreted as Alfv\'en wave
amplitudes, these results allow us to determine the amount of energy per unit
time transferred in the turbulent cascade from large to small scales, and
eventually deposited in the low corona and into the solar wind.  

At distances $r \gapprox 10 \, R_\odot$, the frequency broadening is dominated
by solar wind motion. The deduced velocity values (200-600)~km~s$^{-1}$ at
$\sim$100~$R_\odot$ are consistent with previous scintillation measurements
\citep[e.g.,][]{1971A&A....10..310E,1981A&A...103..415A} and are also consistent
with characteristic solar wind speeds at these distances
\citep[e.g.,][]{2024ApJ...961...64B}. The anisotropy of density fluctuations
appears to be an important ingredient: if the spectrum of density fluctuations
were isotropic, only much slower sub-solar-wind speeds (up to $\sim
100$~km~s$^{-1}$) would be consistent with the frequency broadening
observations; alternatively, the observed frequency broadening would be
consistent with observed solar wind speeds only if the level of density
fluctuations were much lower than inferred from other observations, such as
angular broadening of extra-solar sources. 

Closer to the Sun ($r \lapprox 10 \, R_\odot$), however, the solar wind speed
becomes small, while the velocities required to explain the frequency broadening
observations remain large. The frequency observations require either speeds
(20-70)~km~s$^{-1}$ in the perpendicular direction, or (100-300)~km~s$^{-1}$ in
the parallel direction or both. Within the description adopted, these two scenarios (or
a combination of the two) cannot be meaningfully distinguished.  

Given the possible importance of waves and turbulence in the context of solar
coronal heating, there are a number of reported results on plasma motions in the
corona/solar wind. Plasma motions in the corona between $(1-2) \, R_\odot$ are
normally detected using excess (i.e., larger than what would be from thermal
motion of the emitting ion) broadening of emission lines from minor ions, and
have velocities comparable to those required to realize the observed level of
frequency broadening of radio sources. It should be noted that the non-thermal
broadening is proportional to the line-of-sight speed (i.e. along the $\perp_1$ direction), as distinct from the
speeds inferred from frequency broadening measurements, which are predominantly along the $\perp_2$ direction, 
perpendicular to the line of sight. The similar values of velocity thus suggest azimuthal symmetry 
in the velocity distribution perpendicular to the radial direction, i.e., to the magnetic field.
Similar to the frequency broadening measurements, 
which are agnostic relative to steady flows versus random
``turbulent'' velocity patterns with the same root-mean-square speed, 
there is an ongoing discussion about whether non-thermal broadening 
of spectral lines is due to  bulk plasma flows or to quasi-random 
ion distribution motions \citep[see discussion in][]{2016A&A...590A..99J}. 
Given that ions are preferentially heated
via cyclotron resonance \citep{1997ApJ...484L..87S,1998ApJ...503..475T}, this
distinction could be particularly important.

It is interesting to note that there is a broad agreement among the turbulent
velocities inferred from interplanetary scintillation measurements
\citep[e.g.,][]{1971A&A....10..310E,1981A&A...103..415A}. Our average values are
somewhat smaller, with perpendicular velocities mostly below 100~km~s$^{-1}$,
with a marginal decrease in speed towards the Sun. Importantly, our results,
like those associated with previously reported measurements, show a large spread
of values, re-emphasizing a high level of variability of the turbulence level in
the solar corona.

Scattering of radio waves requires coherent structures (density fluctuations),
which could either be oscillatory in nature or carried by bulk plasma motions.
Perpendicular large-scale motions (at scales much larger than the density
fluctuation wavelength) could be random torsional or kink (e.g., Alfv\'en) waves
that move around small scale fluctuations. In a turbulent plasma, the spectral
broadening may also be associated with large-scale advection of eddies in a
Kolmogorov turbulent cascade \citep[e.g.,][]{1975JFM....67..561T}.
Quasi-parallel motions or waves parallel to the magnetic field with a speed
comparable to the sound (or ion thermal) speed would also produce a similar
frequency broadening. \citet{2019ApJ...871..202W} has interpreted the broadening
as due to sound waves. If ion-sound waves are present, one can calculate the
energy deposited to ions via Landau damping, and we find a value of order
$10^{11}$~erg~g$^{-1}$~s$^{-1}$, comparable to the heating required to sustain a
million-degree corona \citep[e.g.,][]{1986JGR....91.4111H,2005ApJS..156..265C}.

Sound waves do not necessarily propagate from the low atmosphere \citep[although
EUV observations suggest propagating slow/sound waves,
e.g.,][]{1998ApJ...501L.217D}, but could instead be locally generated via
parametric decay of Alfv\'en waves
\citep[e.g.,][]{1969npt..book.....S,1996PhPl....3.4427M,2001A&A...367..705D}
or from MHD turbulence cascade
\citep[e.g.,][]{1993PhFlA...5..257Z,2001ApJ...562..279L,2003MNRAS.345..325C,2009ApJ...707.1668C,2010A&A...519A.114B,2017ApJ...835..147Z}.
Since the value of $\overline {q \, \epsilon^2}$ depends mostly on the level of
density fluctuations near the ion-scale break scale $q_i^{-1}\sim c/\omega_{pi}$
\citep{2023ApJ...956..112K}, the parallel-propagating ion-sound waves Cerenkov
resonate mostly with protons and should be strongly Landau damped. This suggests
that a constant re-supply of ion-sound waves is required, probably via the
aforementioned parametric decay of Alfv\'en waves and/or the turbulent cascade.
Interestingly, the estimate of Kolmogorov cascade power using large scale
motions $v_{\perp}^3/\ell_\perp$ (at the outer scales $\ell_\perp$) provides the
same power that would be dissipated via ion-sound waves at inner scales
$q_i^{-1}$. Within such a scenario, ion-sound waves (or slow MHD mode waves) act
as an intermediate in the coronal heating chain and thus serve as a valuable
diagnostic of ion heating in the solar corona.

\begin{acknowledgments}

We thank the referee for several very helpful suggestions on how to improve the paper. FA (studentship 2604774) and EPK were supported via STFC training grant ST/V506692/1. EPK and DLC were supported via the STFC/UKRI grants ST/T000422/1 and ST/Y001834/1. AGE was supported by NASA Kentucky under NASA award number
80NSSC21M0362 and by the NASA Heliophysics Supporting Research program under
award number 80NSSC24K0244. NC acknowledges funding support from the Initiative Physique des Infinis (IPI), a research training program of the Idex SUPER at Sorbonne Universit\'{e}.

\end{acknowledgments}

\appendix

\section{Diffusion Tensor: Static Density Fluctuations}\label{C:static}

\begin{figure}[!ht]
  \centering
  \includegraphics[width=0.5\textwidth]{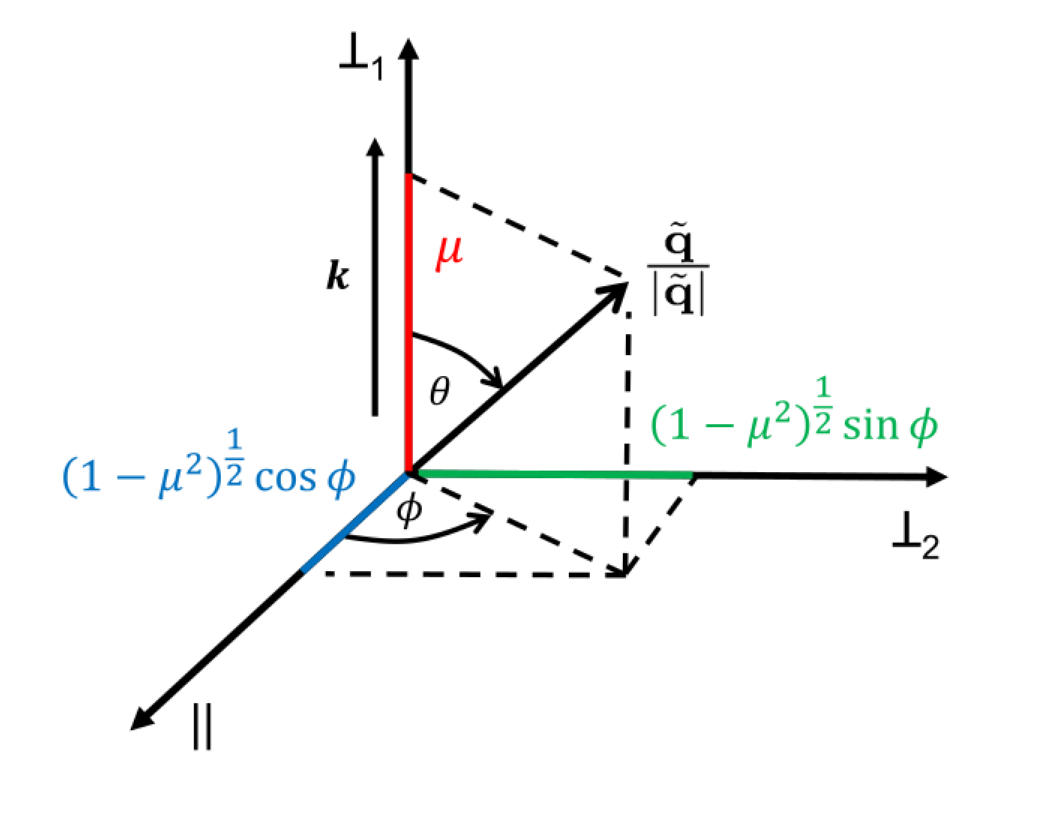}
  \caption{Wavevector coordinate system used in our analysis; the radio wave, with wavevector $\mathbf{k}$, is propagating along the $\perp_1$ direction.}
  \label{fig:coords}
\end{figure}

Figure~\ref{fig:coords} shows the wavevector coordinate system used for our
analysis; it is a polar system in the ${\Tilde {\mathbf q}}$ space (in which the
density fluctuation spectrum $S({\Tilde {\mathbf q}})$ is isotropic: $S({\Tilde
{\mathbf q}}) \equiv S(q)$). The polar axis is aligned with the $\perp_1$
direction, which is along the direction of $\vec{k}$ and $\vec{v}_g$ (Figure~\ref{fig:coords}).
The direction cosine $\mu = \cos \, \theta$, where $\theta$ is the polar angle
from the $\perp_1$ axis, while the azimuthal angle $\phi$ measures the angle
from the $\parallel$ direction in the
$({\Tilde q}_\parallel, {\Tilde q}_{\perp_2})$ plane. Thus ${\Tilde q}_{\perp_1}
= {\Tilde q} \, \mu$, ${\Tilde q}_{\perp_2} = {\Tilde q} \, (1 - \mu^2)^{1/2} \,
\sin \phi$ and ${\Tilde q}_\parallel = {\Tilde q} \, (1 - \mu^2)^{1/2} \, \cos
\phi$. Changing variables from $\vec{q}=(q_\parallel, q_{\perp_2}, q_{\perp_1})$
to $\Tilde{\vec{q}} = \vec{A} \, \vec{q} = ( \alpha^{-1} q_\parallel,
q_{\perp_2}, q_{\perp_1})$, Equation~(\ref{eq:D}) can be written as

\begin{equation}\label{eq:Dig_qtilde}
D_{ij} = \frac{\pi \omega_{p e}^4}{4 \, \omega^2} \, \int q_i \, q_j \, S(\vec{q}) \, \delta\left(\vec{q} \cdot \vec{v}_g\right) \, \frac{d^3 q}{(2 \pi)^3}  = \frac{\pi \omega_{p e}^4}{4 \, \omega^2} \, \alpha \, A_{i \alpha}^{-1} A_{j \beta}^{-1} \int \Tilde{q}_\alpha \, \Tilde{q}_\beta\, S(\vec{\Tilde{q}}) \, \delta\left(\vec{\Tilde{q}} \cdot \vec{\Tilde{v}_g}\right) \, \frac{d^3 \Tilde{q}}{(2 \pi)^3} \,\,\, ,
\end{equation}
where ${\vec {\Tilde v}} = (\alpha \, v_\parallel, v_{\perp_2}, v_{\perp_1})$
and we have used the determinant of the Jacobian $\det{(\textbf{J})} =
\det{(\textbf{A}^{-1})} = \alpha$.

For elastic scattering and a radio wave propagating along the $\perp_1$
direction, we can use Equation~\eqref{eq:Dig_qtilde} to find the various components of the diffusion tensor, viz.

\begin{eqnarray}\label{eq:D_parpar} D_{\parallel \parallel} & = & \frac{\pi \,
\omega_{p e}^4}{4 \, \omega^2} \, \alpha^3 \int_{{\Tilde q} = 0}^\infty
\int_{\phi=0}^{2\pi} \int_{\mu=-1}^1 \Tilde{q}^2 \, (1-\mu^2) \, \cos ^2\phi \,
S({\Tilde{q}}) \, \delta\left(\Tilde{q} \, \Tilde{v}_g \, \mu \right) \,
\frac{\Tilde{q}^2 \, d\Tilde{q} \, d\mu \, d\phi}{(2 \pi)^3} \cr & = &
\frac{\pi^2  \, \omega_{p e}^4}{4 \, \omega^2 \, c} \, \alpha^3 \, \int_{{\Tilde
q} = 0}^\infty \Tilde{q}  \, S({\Tilde{q}}) \, \frac{\Tilde{q}^2 \, d
\Tilde{q}}{(2 \pi)^3} = \frac{\pi  \omega_{p e}^4}{16 \,  \omega^2 \, c} \,
\alpha^2 \, \overline{q \, \eps^2} \,\,\, ,
\end{eqnarray}

\begin{eqnarray}\label{dperp2_perp2_basic} D_{\perp_2 \perp_2} & = & \frac{\pi
\omega_{p e}^4}{4 \, \omega^2} \, \alpha \int_{{\Tilde q} = 0}^\infty
\int_{\phi=0}^{2\pi} \int_{\mu=-1}^1  \Tilde{q}^2 \, (1-\mu^2) \, \sin^2 \phi \,
S({\Tilde{q}}) \, \delta\left(\Tilde{q} \, \Tilde{v}_g \, \mu \right) \,
\frac{\Tilde{q}^2 \, d\Tilde{q} \, d\mu \, d\phi}{(2 \pi)^3} \cr & = & \frac{\pi
\omega_{p e}^4}{4 \, \omega^2 \, c} \, \alpha \int_{{\Tilde q} = 0}^\infty
\int_{\phi=0}^{2\pi}\Tilde{q} \, \sin^2 \phi \, S({\Tilde{q}}) \,
\frac{\Tilde{q}^2 \, d\Tilde{q} \, d\phi}{(2 \pi)^3} =  \frac{\pi \omega_{p
e}^4}{16 \, \omega^2 c} \, \overline{q \, \eps^2} \,\,\, .
\end{eqnarray}
and

\begin{equation}\label{d_perp1_perp1_basic}
D_{\perp_1 \perp_1} = \frac{\pi \omega_{p e}^4}{4 \, \omega^2} \, \alpha \int_{{\Tilde q} = 0}^\infty \int_{\phi=0}^{2\pi} \int_{\mu=-1}^1  \Tilde{q}^2 \, \mu^2 \, S({\Tilde{q}}) \, \delta\left(\Tilde{q} \, \Tilde{v}_g \, \mu \right) \, \frac{\Tilde{q}^2 \, d\Tilde{q} \, d\mu \, d\phi}{(2 \pi)^3} = 0 \,\,\, .
\end{equation}
Further, because $\int _{0} ^{2 \pi} \sin \phi \, \cos \phi \,  d\phi = 0$ we
have $D_{\perp_2 \parallel} = D_{\parallel \perp_2} = 0$, while the Dirac delta
function in the integral results in $D_{\perp_1 \parallel} = D_{\parallel
\perp_1} = D_{\perp_1 \perp_2} = D_{\perp_2 \perp_1} = 0$. Thus

\begin{equation} \label{eq:Delastic}
\mmatrix{D} = \frac{\pi \omega_{p e}^4}{16 \, \omega^2 \, c} \overline{q \, \eps^2} \, \begin{pmatrix}
\alpha^2 & 0 & 0 \\
0 & 1 & 0 \\
0 & 0 & 0
\end{pmatrix} \,\, \equiv \frac{\pi \omega_{p e}^4}{16 \, \omega^2 \, c} \, \overline{q \, \eps^2} \, \, \text{diag}\left(\alpha^2, 1, 0\right) \,\,\, ,
\end{equation}
so that there are only two non-zero elements, neither of which contributes to
frequency broadening. These results recover the expressions obtained by
\citet{2019ApJ...884..122K}.

\section{Diffusion Tensor: Moving Density
Fluctuations}\label{sec:appendix_moving_fluctuations}

\subsection{Parallel Waves}\label{APP:D:1}

Consider density fluctuations moving along the $\parallel$ direction, i.e.,
$\Omega (\vec{q})=v_\parallel \, q_\parallel$. Then from
Equation~\eqref{eq:Dmoving}, and referring to Figure~\ref{fig:coords}, we find

\begin{equation}
D_{\parallel \parallel} =
\frac{\pi \omega_{p e}^4}{4 \, \omega^2} \, \alpha^3 \int_{{\Tilde q} = 0}^\infty \int_{\phi=0}^{2\pi} \int_{\mu=-1}^1 \Tilde{q}^2 \, (1-\mu^2) \cos ^2\phi \, S({\Tilde{q}}) \,
\delta\left(\Tilde{v}_\parallel \, \Tilde{q} \, \sqrt{1-\mu^2} \, \cos\phi - \Tilde{q} \, \Tilde{v}_g \, \mu \right) \, \frac{\Tilde{q}^2 \, d\Tilde{q} \, d\mu \, d\phi}{(2 \pi)^3} \,\,\, ,
\end{equation}

\begin{equation}
D_{\perp_2 \perp_2} = \frac{\pi \omega_{p e}^4}{4 \, \omega^2} \, \alpha \int_{{\Tilde q} = 0}^\infty \int_{\phi=0}^{2\pi} \int_{\mu=-1}^1 \Tilde{q}^2 \, (1-\mu^2) \sin ^2\phi \, S({\Tilde{q}}) \,
\delta\left(\Tilde{v}_\parallel \, \Tilde{q} \, \sqrt{1-\mu^2} \, \cos\phi - \Tilde{q} \, \Tilde{v}_g \, \mu \right) \, \frac{\Tilde{q}^2 \, d\Tilde{q} \, d\mu \, d\phi}{(2 \pi)^3} \,\,\, ,
\end{equation}
and

\begin{equation}
D_{\perp_1 \perp_1} =
\frac{\pi \omega_{p e}^4}{4 \, \omega^2} \, \alpha \,  \int_{{\Tilde q} = 0}^\infty \int_{\phi=0}^{2\pi} \int_{\mu=-1}^1 \Tilde{q}^2 \, \mu^2 \, S({\Tilde{q}}) \, \delta \left (
{\Tilde v}_\parallel \sqrt{1-\mu ^2} \, \Tilde{q} \, \cos\phi - \Tilde{q} \, \Tilde{v}_g \, \mu \right) \, \frac{\Tilde{q}^2 \, d\Tilde{q} \, d\mu \, d\phi}{(2 \pi)^3} \,\,\, .
\end{equation}

Noting that ${\vec {\Tilde v}} = (\alpha \, v_\parallel, v_{\perp_2},
v_{\perp_1})$, the delta function can be expanded using the roots of $g(\mu) = A
\sqrt{1-\mu^2} \, - \mu $, where $A=\left(\tilde{v}_{\|} / \tilde{v}_g\right)
\cos \phi \simeq \alpha \, \left(v_{\|} / c\right) \cos \phi$, into

$$\delta\left( A \sqrt{1-\mu ^2} \, - \mu  \right) = \frac{1}{1 + A^2} \, \delta
 \left( \mu - \frac{A}{\sqrt{1 + A^2}}\right). $$ Thus, integrating over $\mu$
 and retaining up to $\mathcal{O}\left(A^2\right)$ terms, we obtain

\begin{equation}
D_{\parallel \parallel} \simeq \frac{\pi \omega_{p e}^4}{4 \, \omega^2 \, c} \, \alpha^3 \int_{\tilde{q}=0}^{\infty} \int_{\phi=0}^{2 \pi} \tilde{q} \, \left(1-2 \, \frac{{\tilde{v_{\|}}}^2}{c^2} \cos^2 \phi \right) \, \cos ^2 \phi \, S(\tilde{q}) \, \frac{\tilde{q}^2 \, d \tilde{q} \, d \phi}{(2 \pi)^3}=\frac{\pi \omega_{p e}^4}{16 \, \omega^2 \, c} \, \alpha^2 \, \left(1-\frac{3}{2} \, \alpha^2 \, \frac{v_\|^2}{c^2} \right ) \, \overline{q \, \epsilon^2} \,\,\, ,
\end{equation}

\begin{equation}
D_{\perp_2 \perp_2}=\frac{\pi \omega_{p e}^4}{4 \, \omega^2 \, c} \, \alpha \int_{\tilde{q}=0}^{\infty} \int_{\phi=0}^{2 \pi} \tilde{q} \, \left(1-2 \, \frac{{\tilde{v_{\|}}}^2}{c^2} \cos ^2 \phi\right) \, \sin ^2 \phi \, S(\tilde{q}) \, \frac{\tilde{q}^2 \, d \tilde{q} \, d \phi}{(2 \pi)^3} = \frac{\pi \omega_{p e}^4}{16 \, \omega^2 \, c} \, \left( 1-\frac{1}{2} \, \alpha^2 \, \frac{v_\|^2}{c^2} \right) \, \overline{q \, \epsilon^2} \,\,\, ,
\end{equation}
and

\begin{equation}
D_{\perp_1 \perp_1}=\frac{\pi \omega_{p e}^4}{4 \, \omega^2 \, c} \, \alpha \, \int_{\tilde{q}=0}^{\infty} \int_{\phi=0}^{2 \pi} \tilde{q} \, \left(\frac{\tilde{v}_{\|}^2}{c^2} \cos ^2 \phi\right) \, S(\tilde{q}) \, \frac{\tilde{q}^2 \, d \tilde{q} \, d \phi}{(2 \pi)^3} = \alpha^2 \, \frac{v_{\|}^2}{c^2} \, \frac{\pi \omega_{p e}^4}{16 \, \omega^2 \, c} \, \overline{q \, \epsilon^2} \,\,\, .
\end{equation}
These are modified scattering rate expressions, compared to
Equations~\eqref{d_perp1_perp1_basic}, \eqref{dperp2_perp2_basic} and
~\eqref{eq:D_parpar}. However, the corrections are generally small since
$v_\parallel^2/c^2 \ll 1$.

Further, because $\int_{0} ^{2 \pi} \sin \phi \, d\phi = 0$, $\int_{0} ^{2 \pi}
\cos \phi  \, d\phi = 0$, and $\int_{0} ^{2 \pi} \sin \phi \, \cos \phi  \,
d\phi = 0$, we have, respectively,

\begin{equation}
    D_{\perp_1 \perp_2} = D_{\perp_2 \perp_1} = 0 \, ; \qquad D_{\parallel \perp_1} = D_{\perp_1 \parallel} = 0 \, ; \qquad D_{\parallel \perp_2} = D_{\perp_2 \parallel}  = 0 \,\,\, .
\end{equation}
Thus the diffusion tensor has the form

\begin{equation}\label{App_D:diff_tens}
\mmatrix{D}=\frac{\pi \omega_{p e}^4}{16 \, \omega^2 \, c} \,  \, \overline{q \, \epsilon^2} \, \, \text{diag} \left ( \alpha^2 \, \left [ 1 - \frac{3}{2} \, \alpha^2 \, \frac{v_{\|}^2}{c^2} \right ], \,  1-\frac{1}{2} \, \alpha^2 \, \frac{v_{\|}^2}{c^2}, \, \alpha^2 \, \frac{v_{\|}^2}{c^2} \right ) \,\,\, .
\end{equation}

Both $\parallel$ and $\perp_2$ directions are perpendicular to the wavevector
$\mathbf{k}$ in this analysis. Hence, for $\Delta \mathbf{k} \perp \mathbf{k}$
and hence the scattering corresponding to the $\perp_2 \perp_2$ and $\parallel
\parallel$ terms is elastic. The change to the absolute value of $|\mathbf{k}|$,
or equivalently the wave frequency $\omega(\mathbf{k}) \simeq c|\mathbf{k}|$,
comes from the term $D_{\perp_1 \perp_1} \propto d\left\langle\Delta
k_{\perp_1}^2\right\rangle / d t \neq 0$. Henceforth, we can write

\begin{equation}\label{eq:D-parallel-motions}
\mmatrix{D} = \frac{\pi \omega_{p e}^4}{16 \, \omega^2 \, c} \, \, \overline{q \, \epsilon^2} \, \,\text{diag}\left(\alpha^2, 1, \frac{\alpha^2 \, v_{\|}^2}{c^2}\right) \,\,\, .
\end{equation}

\subsection{Perpendicular Waves} \label{APP:D:2}

For the case of waves moving in the $\perp_2$ direction, i.e., $\Omega (\vec{q}) = v_{\perp_2}
q_{\perp_2}$, the same formalism can be applied here, with
$A=(\Tilde{v}_{\perp_2} / \Tilde{v}_g) \, \sin \phi \simeq (v_{\perp_2} / c) \,
\sin \phi$, to obtain

\begin{equation}
D_{\parallel \parallel} \simeq \frac{\pi \omega_{p e}^4}{4 \, \omega^2 \, c } \, \alpha^3 \int_{\tilde{q}=0}^{\infty} \int_{\phi=0}^{2 \pi} \tilde{q} \, \left(1 - 2 \, \frac{{\tilde{v}_{\perp_2}}^2}{c^2} \sin ^2 \phi\right) \, \cos ^2 \phi \, S(\tilde{q}) \, \frac{\tilde{q}^2 \, d \tilde{q} \, d \phi}{(2 \pi)^3}=\frac{\pi \omega_{p e}^4}{16 \, \omega^2 \, c} \, \alpha^2 \, \left(1-\frac{1}{2} \, \frac{{v_{\perp_2}}^2}{c^2}\right) \,\overline{q \, \epsilon^2} \,\,\, ,
\end{equation}

\begin{equation}
D_{\perp_2 \perp_2}=\frac{\pi \omega_{p e}^4}{4 \, \omega^2 \, c } \, \alpha \int_{\tilde{q}=0}^{\infty} \int_{\phi=0}^{2 \pi} \tilde{q} \, \left(1 - 2 \, \frac{{\tilde{v}_{\perp_2}}^2}{c^2} \sin ^2 \phi\right) \, \sin ^2 \phi \, S(\tilde{q}) \, \frac{\tilde{q}^2 \, d \tilde{q} \, d \phi}{(2 \pi)^3}=\frac{\pi \omega_{p e}^4}{16 \, \omega^2 \, c} \, \left(1-\frac{3}{2} \, \frac{{v_{\perp_2}}^2}{c^2}\right) \, \overline{q \, \epsilon^2} \,\,\, ,
\end{equation}
and

\begin{equation}
D_{\perp_1 \perp_1} =
\frac{\pi \omega_{p e}^4}{4 \, \omega^2 \, c } \, \alpha \int_{{\Tilde q} = 0}^\infty \, \int_{\phi = 0}^{2\pi} \Tilde{q} \, \left (\frac{\Tilde{v}_{\perp_2}^2}{c^2} \sin ^2\phi \right ) \, S({\Tilde{q}}) \, \frac{\Tilde{q}^2 \, d\Tilde{q} \, d\phi}{(2 \pi)^3} = \frac{v_{\perp_2}^2}{c^2} \, \frac{\pi  \omega_{p e}^4}{16 \, c\, \omega^2} \, \overline{q \, \eps^2} \,\,\, .
\end{equation}
Again, because $\int_{0} ^{2 \pi} \sin \phi \, d\phi = 0$, $\int_{0} ^{2 \pi}
\cos \phi  \, d\phi = 0$, and $\int_{0} ^{2 \pi} \sin \phi \, \cos \phi  \,
d\phi = 0$, we have, respectively,

\begin{equation}
    D_{\perp_1 \perp_2} = D_{\perp_2 \perp_1} = 0 \, ; \qquad D_{\parallel \perp_1} = D_{\perp_1 \parallel} = 0 \, ; \qquad D_{\parallel \perp_2} = D_{\perp_2 \parallel}  = 0 \,\,\, .
\end{equation}
Thus, following the same reasoning as in Equation (\ref{App_D:diff_tens}), the
diffusion tensor for perpendicular motions takes the form

\begin{equation}\label{eq:D:perp-motions}
\mmatrix{D} = \frac{\pi \omega_{p e}^4}{16 \, \omega^2 \, c} \, \overline{q \, \epsilon^2} \,
\,\text{diag}\left(\alpha^2, 1, \frac{v_{\perp_2}^2}{c^2}\right) \,\,\, .
\end{equation}

\section{Diffusion Tensor: Random Motions} \label{APP_E}

\subsection{Random Motions Superimposed on a Static
Background}\label{D_random_0}

Integrating Equation~\eqref{eq:D_random_plus_flows_general} in the approximation
$q_{\perp_2}^2 \langle v_{\perp_2}^2\rangle \ll {q}^2 c^2$ and taking $\Omega
=0$ yields

\begin{eqnarray}\label{AppD1:random_motion_pp} D_{\parallel \parallel} & = &
\frac{\pi \omega_{p e}^4}{4 \, \omega^2} \, \alpha^3 \int_{{\Tilde q} =
0}^\infty \int_{\phi=0}^{2\pi} \int_{\mu=-1}^1 \Tilde q^2 \, (1 - \mu^2) \,
\cos^2 \phi \, S(\Tilde q) \, \frac{1}{\sqrt{2 \pi \,\Tilde q_{\perp_2}^2
\langle \Tilde v_{\perp_2}^2\rangle}} \, \exp \left[- \, \frac{( \Tilde q \, c
\, \mu)^2}{2 \, \Tilde q_{\perp_2}^2 \langle \Tilde
v_{\perp_2}^2\rangle}\right]\, \frac{d\mu \, d\phi \,\Tilde q^2 d \Tilde q \,
}{(2 \pi)^3} \, \cr & = &  \frac{\pi \omega_{p e}^4}{16 \, \omega^2 \, c} \,
\alpha^2 \,\left( 1- \frac{1}{4} \frac{\langle v_{\perp_2}^2 \rangle}{c^2}
\right)\, \overline{q \, \eps^2} \,\,\, ,
\end{eqnarray}
where we have used the substitution $\xi = \mu \, c/\sqrt{\langle \Tilde
v_{\perp_2}^2\rangle}$, approximated $ q_{\perp_2}^2 \approx q^2 \sin^2 \chi$
and made use of the formula  $\int_{-\infty}^{\infty} \xi^2 \exp{(-b \, \xi^2)}
\, d \xi= \sqrt{\pi/4 \, b^3}$. In the same way, we find

\begin{eqnarray}\label{AppD1:random_motion_22} D_{\perp_2\perp_2} & = &
\frac{\pi \omega_{p e}^4}{4 \, \omega^2} \, \alpha \int_{{\Tilde q} = 0}^\infty
\int_{\phi=0}^{2\pi} \int_{\mu=-1}^1 \Tilde q^2 \, (1 - \mu^2) \, \sin^2 \phi \,
S(\Tilde q) \, \frac{1}{\sqrt{2 \pi \,\Tilde q_{\perp_2}^2 \langle \Tilde
v_{\perp_2}^2\rangle}} \, \exp \left[- \, \frac{( \Tilde q \, c \, \mu)^2}{2 \,
\Tilde q_{\perp_2}^2 \langle \Tilde v_{\perp_2}^2\rangle}\right]\, \frac{d\mu \,
d\phi \,\Tilde q^2 d \Tilde q \, }{(2 \pi)^3} \, \cr & = &  \frac{\pi \omega_{p
e}^4}{16 \, \omega^2 \, c} \,\left( 1-\frac{3}{4} \frac{\langle v_{\perp_2}^2
\rangle}{c^2} \right ) \, \overline{q \, \eps^2} \,\,\, ,
\end{eqnarray}

\begin{eqnarray}\label{AppD1:random_motion_11} D_{\perp_1\perp_1} & = &
\frac{\pi \omega_{p e}^4}{4 \, \omega^2} \, \alpha \int_{{\Tilde q} = 0}^\infty
\int_{\phi=0}^{2\pi} \int_{\mu=-1}^1 \Tilde q^2 \, \mu^2 \, S(\Tilde q) \,
\frac{1}{\sqrt{2 \pi \,\Tilde q_{\perp_2}^2 \langle \Tilde
v_{\perp_2}^2\rangle}} \, \exp \left[- \, \frac{( \Tilde q \, c \, \mu)^2}{2 \,
\Tilde q_{\perp_2}^2 \langle \Tilde v_{\perp_2}^2\rangle}\right] \, \frac{d\mu
\, d\phi \,\Tilde q^2 d \Tilde q \, }{(2 \pi)^3} \, \cr & = & \frac{\langle
v_{\perp_2}^2\rangle}{c^2} \frac{\pi \omega_{p e}^4}{16 \, \omega^2 \, c} \,
\overline{q \, \eps^2} \,\,\, ,
\end{eqnarray}
and 
\begin{equation}
    D_{\perp_1 \perp_2} = D_{\perp_2 \perp_1} = 0 \, ; \qquad D_{\parallel \perp_1} = D_{\perp_1 \parallel} = 0 \, ; \qquad D_{\parallel \perp_2} = D_{\perp_2 \parallel}  = 0 \,\,\, .
\end{equation}
The components of the diffusion tensor for random motions (``turbulence'') can
then be approximated as

\begin{equation}\label{eq:random-static}
\mmatrix{D} =\frac{\pi \omega_{p e}^4}{16 \, \omega^2 \, c} \, \overline{q \, \eps^2} \,
\,\text{diag} \left( \alpha^2, 1, \frac{\langle v_{\perp_2}^2 \rangle}{c^2} \right ) \,\,\, .
\end{equation}

\subsection{Random Motions Superimposed on Flows in a General
Direction}\label{D_random_1}

Here we consider random motions superimposed on flows in a general direction perpendicular to $\mathbf{k}$.
For this purpose we can use the general result

\[
\int_{\xi = -\infty}^{\infty} B \, \xi^2 \, \exp{\left(- \frac{B^2}{2}(\xi - A)^2\right)} \, d \xi =
B \, \int_{\eta = -\infty}^{\infty} (A+\eta)^2 \, \exp{ \left (- \frac{B^2}{2} \eta^2 \right ) } \, d \eta =
\]

\begin{equation}\label{eq:random-parallel}
= B \, \int_{\eta = -\infty}^{\infty} (A^2 + \eta^2 + 2A\eta ) \, \exp{ \left ( - \frac{B^2}{2} \eta^2 \right )} \, d \eta =  \sqrt{2\pi} \left ( A^2 + \frac{1}{B^2} \right )
\end{equation}
to compute the components of the diffusion tensor.

We first consider radially propagating density fluctuations, with $\Omega
(\vec{q})=v_\parallel \, q_\parallel$. If we add random motions in the same
direction we can evaluate the $D_{\perp_1\perp_1}$ term that contributes to the
frequency broadening:

\begin{eqnarray}\label{AppD2:random_motion_11} D_{\perp_1\perp_1} & = &
\frac{\pi \omega_{p e}^4}{4 \, \omega^2} \, \alpha \int_{{\Tilde q} = 0}^\infty
\int_{\phi=0}^{2\pi} \int_{\mu=-1}^1 \Tilde q^2 \, \mu^2 \, S(\Tilde q) \,
\frac{1}{\sqrt{2 \pi \, \Tilde q_{\parallel}^2 \langle \Tilde
v_{\parallel}^2\rangle}} \, \exp \left[- \, \frac{( \Tilde q \, c \, \mu -
\Tilde q_\parallel \Tilde v_\parallel )^2}{2 \, \Tilde q_{\parallel}^2 \langle
\Tilde v_{\parallel}^2\rangle}\right]\, \frac{d\mu \, d\phi \,\Tilde q^2 d
\Tilde q \, }{(2 \pi)^3} \, \cr
& = & \frac{\pi \omega_{p e}^4}{16 \, \omega^2 \, c} \, \overline{q \, \eps^2}
\, \frac{ \alpha^2 ( v_\parallel^2 + \langle v_{\parallel}^2\rangle)}{c^2}
\,\,\, ,
\end{eqnarray}
where we have approximated $\Tilde q_{\parallel}^2  \approx \Tilde q^2 \cos^2
\phi$, made use of the substitution $\mu = \xi \sqrt{\langle {\Tilde
v}_{\parallel}^2\rangle} \, / \, c$, and used
Equation~\eqref{eq:random-parallel}. 

If we then consider propagation and random motions that are both in the
direction perpendicular to the solar radius vector, with dispersion relation
$\Omega (\vec{q})=v_{\perp_2} \, q_{\perp_2}$, we find

\begin{eqnarray}\label{AppD2:random_motion_22} D_{\perp_1\perp_1} & = &
\frac{\pi \omega_{p e}^4}{4 \, \omega^2} \, \alpha \int_{{\Tilde q} = 0}^\infty
\int_{\phi=0}^{2\pi} \int_{\mu=-1}^1 \Tilde q^2 \, \mu^2 \, S(\Tilde q) \,
\frac{1}{\sqrt{2 \pi \, \Tilde q_{\perp_2}^2 \langle \Tilde
v_{\perp_2}^2\rangle}} \, \exp \left[- \, \frac{( \Tilde q \, c \, \mu - \Tilde
q_{\perp_2} \Tilde v_{\perp_2} )^2}{2 \, \Tilde q_{\perp_2}^2 \langle \Tilde
v_{\perp_2}^2\rangle}\right]\, \frac{d\mu \, d\phi \,\Tilde q^2 d \Tilde q \,
}{(2 \pi)^3} \, \cr & = & \frac{\pi \omega_{p e}^4}{16 \, \omega^2 \, c} \,
\overline{q \, \eps^2} \, \frac{ v_{\perp_2}^2 + \langle
v_{\perp_2}^2\rangle}{c^2} \,\,\, ,
\end{eqnarray}
where we have made use of the substitution $\mu = \xi \sqrt{\langle {\Tilde
v}_{\perp_2}^2\rangle} \, / \, c$, approximated $\Tilde q_{\perp_2}^2  \approx
\Tilde q^2 \sin^2 \chi$, and used Equation~\eqref{eq:random-parallel}.

If both parallel and perpendicular contributions are taken into account, the
diffusion tensor takes the form

\begin{equation}\label{eq:random-radial}
\mmatrix{D} =\frac{\pi \omega_{p e}^4}{16 \, \omega^2 \, c} \, \overline{q \, \eps^2} \,
\,\text{diag} \left( \alpha^2, 1, \frac{\alpha^2 \left( v_\parallel^2 + \langle
 v_{\parallel}^2\rangle\right) + v_{\perp_2}^2 + \langle
v_{\perp_2}^2\rangle }{c^2} \right ) \,\,\, .
\end{equation}
Since steady flows $v^2$ and random motions $\langle  v^2\rangle$ contribute
equally, we can rewrite this as simply

\begin{equation}\label{eq:random-radial_v_par_perp}
\mmatrix{D} =\frac{\pi \omega_{p e}^4}{16 \, \omega^2 \, c} \, \overline{q \, \eps^2} \,
\,\text{diag} \left( \alpha^2, 1, \frac{\alpha^2 \, \langle
 v_{\parallel}^2\rangle + \langle
v_{\perp_2}^2\rangle}{c^2} \right ) \,\,\, ,
\end{equation}
where now $\langle \cdots \rangle$ includes both steady and random flows, in the
parallel or perpendicular directions, respectively, added in quadrature.

\section{Conversion of Reported Frequency Broadening Values to Standard
Deviations}\label{sec:conversion-to-sigma}

Here we briefly summarize how the frequency broadening in each reported data
set, that is not already presented as a standard deviation $\sigma$, is
converted to give $\Delta{f} \equiv \sigma$ for use in
Figure~\ref{fig:freq_broadening_obs}. We also note that any reported data that
relate to solar transient events have been removed, and we consider only one-way
signals.

\cite{goldstein1967superior} define the bandwidth as the width of an equivalent
rectangle of the same height and area as the measured curves. Comparing with a
normalized Gaussian distribution, this implies that the reported bandwidth is $B
= \sqrt{2 \pi} \, \sigma$, so that $\Delta{f} = \sigma= B/\sqrt{2\pi}$.
\cite{1980SvA....24..454Y} define the bandwidth as the ``width of the spectral
line,'' which, absent more detailed specification, we take to be a measure of
the standard deviation $\sigma$. The signal measurements of
\cite{2003ITAP...51..201M} are provided as $B$ = FWHM, which converts as
$\Delta{f}=\sigma=B/2\sqrt{2\ln{2}}$. Finally, \cite{1976ApJ...210..593W,
1978ApJ...219..727W, 1979JGR....84.7288W} and \cite{1980MNRAS.190P..73B} define
the bandwidth $B$ through the relation

\begin{equation}\nonumber
    \int_0^{B/2} P(f) \; \mathrm{d}f = \frac{1}{2} \int_0^\infty P(f) \;\mathrm{d}f \,\,\, .
\end{equation}
For a Gaussian distribution with standard deviation $\sigma$, this reduces to

\begin{equation}\nonumber
    {\rm erf} \left ( \frac{B}{2 \, \sqrt{2} \, \sigma } \right )= \frac{1}{2} \,\,\, ,
\end{equation}
where the error function is ${\rm erf}(x) = \int_0^x e^{-t^2} \, dt$. We thus
obtain, for these data sets, 

\begin{equation}\nonumber
    \Delta{f} = \sigma = \left [ \, 2\sqrt{2} \, {\rm erf}^{-1} \left ( \frac{1}{2} \right ) \, \right ]^{-1} B  \simeq 0.75 \, B \,\,\, .
\end{equation}

\section{Non-thermal Velocities}\label{sec:non-thermal_vel}

The non-thermal velocities are determined from the width $\zeta$ of the spectrum
line profile, measured at the $1/e$ level, and we retrieve the standard
deviation $v_{\mathrm{nth}}=\zeta/\sqrt{2}$. We use data from
\cite{1990ApJ...348L..77H, 1991SoPh..131...25C, 1998A&A...339..208B,
1998SoPh..181...91D, 1999A&A...349..956D, 1999ApJ...510L..63E,
2004AnGeo..22.3055C, 2009A&A...501L..15B, 2009ApJ...691..794L,
2011SoPh..270..213S}; and \cite{2012ApJ...751..110B}. A summary of the results
is presented in Figure~\ref{fig:df2velEff}.

\section{Analytical approximation to frequency
broadening}\label{sec:analytic_df}

For the propagation of a radio wave from a distant radio source to the observer
at the Earth with a heliocentric angular separation $r_{\rm obs} = r(z=0)$
\citep[see, e.g.,][and Figure~\ref{fig:point_source_size}]{2023ApJ...956..112K},
there is a frequency broadening given by Equation~\eqref{eq:int_los_df}:

\begin{equation}\label{eq:frequency_boadening_random_static}
  \frac{\langle \Delta \, \omega ^2\rangle}{\omega ^2 } = \frac{\pi}{8 \, c^2 \, \omega^4} \, \int _{los} \, \omega_{pe}^4 \left [
   \alpha ^2 \, \langle v_\parallel^2 \rangle \cos^2 \chi \, + \langle v_{\perp}^2 \rangle (1+\sin^2 \chi) \,  \right ] \, \overline{q \, \eps ^2}  \,\, dz \,\,\, .
\end{equation}
Assuming that the various quantities in the integral are functions of
heliocentric distance $r = \sqrt{r_{\rm obs}^2 + z^2}$, we obtain

\begin{eqnarray}\label{eq:freq-broadening-expression}
      \frac{\langle \Delta \, \omega ^2\rangle}{\omega ^2 } 
      & = & \frac{2 \, \pi^3 \, e^4}{m_e^2 \, c^2 \, \omega^4} \,
      \int_{-\infty}^{1~{\rm au}}  \, n^2 \left ( \sqrt{r_{\text{obs}}^2+z^2}
      \right ) \,\, \overline{q \, \eps^2} \left ( \sqrt{r_{\text{obs}}^2+z^2}
      \, \right ) \,\,\, \times \cr & \times & \,\,\, \left [ \alpha^2 \,
      \langle v_\parallel^2 \rangle \left ( \sqrt{r_{\text{obs}}^2+z^2} \right )
      \cos^2 \chi \, + \langle v_{\perp}^2 \rangle \left (
      \sqrt{r_{\text{obs}}^2+z^2} \right ) (1+\sin^2 \chi) \, \right ]\, dz
      \,\,\, ,
\end{eqnarray}
where we have used $\omega_{pe} = \sqrt{4 \pi n e^2/m_e}$. With the substitution
$z = r_{\text{obs}} \tan \chi$, this can be written

\begin{eqnarray}\label{eq:freq-broad-1}
      \frac{\langle \Delta \, \omega ^2\rangle}{\omega ^2 }
      & = & \frac{2 \, \pi^3 \, e^4}{m_e^2 \, c^2 \, \omega^4} \,
      \frac{r_{\text{obs}}}{R_\odot} \, \int_{-\pi/2}^{\tan^{-1}(215 \,
      R_\odot/r_{\text{obs}})} \, n^2 ( r_{\text{obs}} \sec \chi ) \,\,
      \overline{q \, \eps^2} \, R_\odot ( r_{\text{obs}} \sec \chi ) \,\,\,
      \times \cr & \times & \,\,\, \left [ \alpha^2 \, \langle v_\parallel^2 \,
      \rangle ( r_{\text{obs}} \sec \chi  ) \cos^2 \chi \, + \, \langle
      v_{\perp}^2 \, \rangle ( r_{\text{obs}} \sec \chi  ) (1+\sin^2 \chi) \,
      \right ] \, \sec^2 \chi \, d\chi \,\,\, .
\end{eqnarray}
Figure~6 of \cite{2023ApJ...956..112K} shows that, empirically,

\begin{equation}\label{eq:n-sq-qeps2-fit}
  n^2 \, ( r_{\text{obs}} \sec \chi ) \,\, \overline{q \, \eps^2} \, R_\odot \, ( r_{\text{obs}} \sec \chi ) \simeq 6.5 \times 10^{14} \, \left ( \frac{r_{\text{obs}} \, \sec \chi}{R_\odot} - 1 \right )^{-5.17}  \, {\rm cm}^{-6} \,\,\, .
\end{equation}
Using this expression and taking the velocity variances outside the integral as
averages, we obtain

\begin{equation}\label{eq:freq-broad-2}
\begin{aligned}
\frac{\left\langle\Delta \omega^2\right\rangle}{\omega^2}= & 6.5 \times 10^{14}\left(\frac{e^4}{8 \pi m_e^2 c^2 f^4}\right)\left(\frac{r_{\mathrm{obs}}}{R_{\odot}}\right)^{-4.17} \times \\
& \times\left\{\alpha^2 \overline{\langle v_{\|}^2\rangle} \int_{-\pi / 2}^{\tan ^{-1}\left(215 R_{\odot} / r_{\mathrm{obs}}\right)}\left(1-\frac{R_{\odot} \cos \chi}{r_{\mathrm{obs}}}\right)^{-5.17} \cos ^{5.17} \chi d \chi+\right. \\
+ & \left.\overline{\left\langle v_{\perp}^2\right\rangle} \int_{-\pi / 2}^{\tan ^{-1}\left(215 R_{\odot} / r_{\mathrm{obs}}\right)}\left(1-\frac{R_{\odot} \cos \chi}{r_{\mathrm{obs}}}\right)^{-5.17} \cos ^{3.17} \chi\left(1+\sin ^2 \chi\right) d \chi\right\} \,\,\, .
\end{aligned}
\end{equation}
At closest-approach distances $r_{\rm obs} \gg R_\odot$, the term $(1 - R_\odot
\cos \chi/r_{\rm obs})^{-5.7} \simeq 1$. Adopting this approximation, the
frequency broadening reduces to the relatively simple form

\begin{equation}\label{eq:freq-broad-3}
\begin{aligned}
\frac{\left\langle\Delta \omega^2\right\rangle}{\omega^2} & =6.5 \times 10^{14}\left(\frac{e^4}{8 \pi m_e^2 c^2 f^4}\right)\left(\frac{r_{\mathrm{obs}}}{R_{\odot}}\right)^{-4.17} \times \\
& \times\left\{\alpha^2 \overline{\langle v_{\|}^2\rangle} \int_{-\pi / 2}^{\chi_{\mathrm{obs}}} \cos ^{5.17} \chi d \chi+\overline{\left\langle v_{\perp}^2\right\rangle} \int_{-\pi / 2}^{\chi_{\mathrm{obs}}} \cos ^{3.17} \chi\left(1+\sin ^2 \chi\right) d \chi\right\} \,\,\, .
\end{aligned}
\end{equation}
where $\chi_{\text {obs }}=\tan ^{-1}\left(215 R_{\odot} / r_{\text {obs
}}\right)$. Each integral can be split into two parts: one from $\chi=-\pi / 2$
to 0 (corresponding to the incoming ray from $\infty$ to the distance of closest
approach $r_{\text {obs }}$, and other from 0 to $\chi_{\text {obs }}$,
corresponding to the outgoing ray from $r_{\text {obs }}$ to $1 \mathrm{au}$.
The integral can then be expressed in terms of beta functions and incomplete
beta functions, respectively, viz.

\begin{equation}\label{eq:freq-broad-final}
\frac{\left\langle\Delta \omega^2\right\rangle^{1 / 2}}{\omega}=2.55 \times 10^7\left(\frac{e^2}{m_e c}\right)\left(\frac{r_{\mathrm{obs}}}{R_{\odot}}\right)^{-2.085}\left[\alpha^2 \, \beta_{\|}^2 \, v_{\|, \mathrm{rms}}^2 + \beta_{\perp}^2 \, v_{\perp, \mathrm{rms}}^2\right]^{1 / 2} \times \frac{1}{f^2} \,\,\, ,
\end{equation}
where we have defined $v_{\parallel,{\rm rms}} = \overline{\langle v_\parallel^2
\, \rangle}^{1/2}$ and $v_{\perp,{\rm rms}} = \overline{\langle v_{\perp}^2 \,
\rangle}^{1/2}$ and 

\begin{equation}\label{eq:beta-functions1}
  \beta_{\|}\left(\frac{r_{\mathrm{obs}}}{R_{\odot}}\right)=\left[\frac{\mathrm{B}(1;3.085,0.5)+\mathrm{B}(\psi;3.085,0.5)}{16 \pi}
  \right]^{1/2} \,\,\, ,
\end{equation}

\begin{equation}\label{eq:beta-functions2}
  \beta_{\perp}\left(\frac{r_{\mathrm{obs}}}{R_{\odot}}\right)=
  \left[\frac{\mathrm{B}(1;2.085,0.5)+\mathrm{B}(\psi;2.085,0.5)+\mathrm{B}(1;2.085,1.5)+\mathrm{B}(\psi;2.085,1.5)}{16 \pi} \right]^{1/2} \,\,\, .
\end{equation}
Here $\mathrm{B}(\psi ; u, v)$ are the (incomplete for $\psi<1$ ) beta functions
corresponding to the integrals $2 \int_0^{\pi / 2} \cos ^{5.17} \chi d \chi$, $
2 \int_0^{\pi / 2} \cos ^{3.17} \chi d \chi$, and $2 \int_0^{\pi / 2} \cos
^{3.17} \chi \sin ^2 \chi d \chi$,  and $\psi=$
$\left[1+\left(r_{\text {obs }} / 215 R_{\odot}\right)^2\right]^{-1}$.

We note that the ratio of the third and first terms in the numerator of the
expression for $\beta_\perp$ is

\[
\rho_\perp = \frac{\Gamma (2.085) \,\, \Gamma (1.5) \, / \, \Gamma (3.585)}{\Gamma (2.085) \,\, \Gamma (0.5) \, / \, \Gamma (2.585)} = \frac{0.5}{2.585} \simeq 0.2 \,\,\, ,
\]
so that, given the $\sqrt{1 + \rho_\perp}$ dependence in the
expression~\eqref{eq:beta-functions1} for $\beta_\perp$, the contribution from
the $\sin^2 \chi$ term (which is associated with motions in the $\perp_1$
direction) to the overall broadening is only about 10\%. Moreover, the ratio of
the lead terms in the expressions for $\beta_\parallel$ and $\beta_\perp$ is

\[
\sqrt \frac{\Gamma (3.085) \,\, \Gamma (0.5) \, / \, \Gamma (3.585)}{\Gamma (2.085) \,\, \Gamma (0.5) \, / \, \Gamma (2.585)} = \sqrt { \frac{2.085}{2.585}} \simeq 0.9 \,\,\, ,
\]
showing that the contributions from motions in the $\parallel$ and $\perp_2$
directions are similar (apart from the anisotropy factor $\alpha$). But, since
$\alpha^2 \ll 1$, we can, to a good approximation, neglect the contribution from
$v_{\parallel, {\rm rms}}$ and write Equation~\eqref{eq:freq-broad-final} as

\begin{equation}\label{eq:freq-broad-good-approximation}
      \frac{\langle \Delta \, \omega ^2\rangle^{1/2}}{\omega} = 2.55 \times 10^7 \,  \beta_\perp \, \left ( \frac{e^2}{m_e \, c} \right ) \, \left ( \frac{r_{\rm obs}} {R_{\odot}}\right )^{-2.085} \, v_{\perp,{\rm rms}}  \,  \times \frac{1}{f^2} \,\,\, .
\end{equation}
Scaling to a nominal frequency of $f = 1$~GHz
(Figure~\ref{fig:freq_broadening_obs}), this evaluates to

\begin{equation}\label{eq:freq-broad-numerical}
      \frac{\Delta f}{f} \simeq 2.1 \times 10^{-13} \, \, \beta_\perp \, v_{\perp,{\rm rms}}  \, \left ( \frac{r_{\text{obs}}}{R_\odot} \right )^{-2.085} \left(\frac{1 \, \text{GHz}}{f}\right)^2  \,\,\,  ,
\end{equation}
where we have written $\Delta f$ for $\langle \Delta \, f ^2\rangle^{1/2}$.
Equation~\eqref{eq:freq-broad-numerical} provides a simple, but nevertheless
accurate, analytical approximation for the frequency broadening, valid for
$r_{\text{obs}} \gg R_\odot$.  With a nominal $r_{\text{obs}}=10 \, R_\odot$, we
obtain $\beta_\perp \simeq 0.25$ and so $\Delta f/f \simeq 4~\times~10^{-16} \,
v_{\perp,{\rm rms}} \, (f[{\rm GHz}])^{-2}$, corresponding to $\Delta f \simeq
4~\times~10^{-7} \, v_{\perp,{\rm rms}}$~Hz at $f = 1 \, {\rm GHz}$.
Figure~\ref{fig:freq_broadening_obs} shows that $\Delta f \simeq 3$~Hz at $r =
10 \, R_\odot$, corresponding to  $v_{\perp,{\rm rms}} \simeq 7.5 \times
10^6$~cm~s$^{-1}$, i.e., 75~km~s$^{-1}$.


\bibliography{refs}

\end{document}